\documentclass[12pt,letterpaper]{JHEP3}

\usepackage{amscd,amsmath,amssymb,amsfonts,xspace,mathrsfs}
\usepackage[utf8]{inputenc}

\newcommand{\be}{\begin{equation}}
\newcommand{\ee}{\end{equation}} 
\newcommand{\bes}{\begin{equation*}}
\newcommand{\ees}{\end{equation*}}
\newcommand{\sgn}{\mathrm{sgn}}

\newcommand{\cD}{\mathcal{D}}
  
\newcommand{\CF}{\mathcal{F}}

\newcommand{\CK}{\mathcal{K}}
 
\newcommand{\CM}{\mathcal{M}} 
\newcommand{\CN}{\mathcal{N}}
\newcommand{\CO}{\mathcal{O}} 

\newcommand{\CQ}{\mathcal{Q}}

\newcommand{\CU}{\mathcal{U}}

\newcommand{\CZ}{\mathcal{Z}}

\newcommand{\BF}{\mathbb{F}}
\newcommand{\BR}{\mathbb{R}}
\newcommand{\BC}{\mathbb{C}}
\newcommand{\BH}{\mathbb{H}}
\newcommand{\BP}{\mathbb{P}}
\newcommand{\BZ}{\mathbb{Z}}
\newcommand{\BQ}{\mathbb{Q}}

\newcommand{\eps}{\epsilon}

\newcommand{\bfb}{{\boldsymbol b}}
\newcommand{\bfC}{{\boldsymbol C}}
\newcommand{\bff}{{\boldsymbol f}} 
\newcommand{\bfm}{{\boldsymbol m}}

\newcommand{\bfx}{{\boldsymbol x}}
\newcommand{\bfy}{{\boldsymbol y}}
\newcommand{\bfrho}{{\boldsymbol \rho}}
\newcommand{\bfnu}{{\boldsymbol \nu}}

\newcommand{\bfk}{{\boldsymbol k}}
\newcommand{\bfz}{{\boldsymbol z}}
\newcommand{\bfmu}{{\boldsymbol \mu}}

\newcommand{\non}{\nonumber}

\title{Donaldson-Witten theory and indefinite theta functions 
}
\author{Georgios Korpas and Jan Manschot\\
{\it School of Mathematics, Trinity College, College Green, Dublin 2, Ireland}\\
{\it Hamilton Mathematical Institute, Trinity College, College Green, Dublin 2, Ireland}\\

\vspace*{2mm} {\tt e-mail:
\email{george.korpas@maths.tcd.ie}, 
\email{manschot@maths.tcd.ie}}
}

\abstract{We consider partition functions with insertions of surface operators of
topologically twisted $\CN=2$, SU(2) supersymmetric Yang-Mills theory,
or Donaldson-Witten theory for short, on a four-manifold. If the metric of the compact four-manifold has positive scalar
curvature, Moore and Witten have shown that the partition function is completely determined by
the integral over the Coulomb branch parameter $a$, while more generally the
Coulomb branch integral captures the wall-crossing behavior of both
Donaldson polynomials and Seiberg-Witten invariants. We show that 
after addition of a $\mathcal{\bar Q}$-exact surface
operator to the Moore-Witten integrand, the integrand can be written as a total derivative to the anti-holomorphic
coordinate $\bar a$ using Zwegers' indefinite theta 
functions. In this way, we reproduce G\"ottsche's expressions for Donaldson invariants of rational surfaces
in terms of indefinite theta functions for any choice of metric.  
}

\preprint{TCD-MATH 17--15}

\begin{document}

\section{Introduction}   
Supersymmetric gauge theories with $\CN=2$ and $\CN=4$ supersymmetry
in four dimensions have been a useful and
powerful tool for advances in both physics and
mathematics. One of the major successes of such quantum field theories
is that their path integral (or partition function) may be exactly
evaluated. See for a small sample of the large literature
\cite{Vafa:1994tf, Moore:1997pc, LoNeSha, Ne, Pestun:2007rz, Kapustin:2009kz, 
  Bershtein:2015xfa}. A seminal result of Witten  
\cite{Witten:1988ze} prior to these remarkable developments, is the construction
of topological quantum field theories by topological twisting of a supersymmetric field theory. Once evaluated on a smooth compact four-manifold $M$ these theories
become of cohomological nature and provide topological invariants of
$M$. These topological invariants are typically based on moduli spaces
of instantons \cite{DONALDSON1990257,Donaldson90}, which are solution spaces to the anti-self-duality
equations $*F=\pm F$ of the Yang-Mills field strength $F$.  
    
We consider in this article the path integral of topologically twisted
$\CN=2$, SU$(2)$ and $\operatorname{SO}(3)$ Yang-Mills theory on a compact four-manifold $M$ without boundary,
the Donaldson-Witten theory for short
\cite{Witten:1988ze}. The full path integral $\CZ_{\text{DW}}$ of this theory can be decomposed in to
a continuous integral $\CZ_{u}$ over the Coulomb branch of the theory, where the gauge 
group is spontaneously broken to U(1) by a non-vanishing
expectation value of the order parameter $u=\frac{1}{16\pi^2}\left<\mathrm{Tr}\,\phi^2\right>$, and
a contribution\footnote{We use the subscript ``SW'',
  since this contribution is fully determined by the so-called
  Seiberg-Witten invariants of $M$ \cite{Witten:1994cg}.} $\CZ_{\rm SW}$ from the points where the
effective theory on the Coulomb branch becomes singular
\cite{Moore:1997pc, LoNeSha, Witten:1994cg, Witten:1995gf}. Schematically, we
then have  
\be 
\CZ_{\rm DW}= \CZ_{\rm SW}+\CZ_{u}.   
\ee
The $u$-plane integral $\CZ_u$ only contributes for manifolds with 
$b_2^+=1$. For this class of
four-manifolds, the path
integral is not quite a topological invariant, but only piecewise
constant as a function of the metric. The metric dependence is
captured by the integral over the Coulomb branch $\CZ_u$. 
 
More accurately, the partition function of Donaldson-Witten theory vanishes without
the insertion of additional operators. Since the theory is topological, these operators
correspond to (co)homology classes of $M$. To this end, assume $M$ is
simply connected and let $p\in H_0(M,\BQ)\cong \BQ$ and
$\bfx\in H_2(M,\BQ)$. Then one can consider the following correlation function 
\be
\label{DWcorr}
\left< e^{p\,\CO^{(0)}+ \int_{\bfx}\CO^{(2)}} \,\right>,  
\ee
where $\CO^{(0)}$ and $\CO^{(2)}$ are a point and surface 
operator, which are discussed in more detail in Section \ref{sec_top_twist}.
A famous aspect of Donaldson-Witten theory is that the correlation
function (\ref{DWcorr}) is a generating function of Donaldson
polynomials defined in mathematics \cite{DONALDSON1990257,Donaldson90}, where they play an important
role in the classification of smooth four-manifolds. The Donaldson
polynomials $P_{c_1,k}(p,\bfx) \in \BQ[p,\bfx]$ are polynomials on the rational homology of $M$,
$P_{c_1,k}: H_*(M,\mathbb{Q})\to
\mathbb{Q}$ and are defined using the geometry of the moduli space
$\mathcal{M}_{c_1,k}$ of $k$-instantons with first Chern class  
$c_1(E)=\frac{i}{2\pi}\mathrm{Tr}\, F$.\footnote{In fact, $P_{c_1,k}(p,\bfx)$
only depends on the second Stiefel-Whitney
class $w_2=c_1 \mod H^2(M,2\BZ)$ of the associated vector bundle.} For the
mathematical definition of Donaldson invariants, one considers the map
$\mu_D$, which maps a cycle $\beta\in H_i(M,\BQ)$ to a cocycle $\mu_D(\beta)\in H^{4-i}(\mathcal{M}_{c_1,k},\BQ) $, that is a ($4-i$)-form
on $\mathcal{M}_{c_1,k}$ of the corresponding vector bundle. Restricting to a simply connected closed
four-manifold $M$ and letting $p\in H_0(M,\BQ)$ and $\bfx\in
H_2(M,\BQ)$, the Donaldson
polynomial is defined as the following integral over $\mathcal{M}_{c_1,k}$,
\be
\label{Don_pol}
\begin{split}
P_{c_1,k}(p,\bfx)&=\sum_{r,s\geq 0} \int_{\mathcal{M}_{c_1,k}}
\mu_D(p)^r\wedge\mu_D(\bfx)^s.
\end{split}
\ee 
The integral gives a non-vanishing result only if the degree of the
integrand matches the dimension of $\CM_{c_1,k}$, therefore
$2r+s=\dim_{\BC}\mathcal{M}_{c_1,k}$ and (\ref{Don_pol}) is
indeed a polynomial. The correlation function (\ref{DWcorr}) equals 
the sum of (\ref{Don_pol}) over $k$ \cite{Witten:1988ze}
\be
\Phi_{\bfmu}(p,\bfx)=\sum_k P_{c_1,k}(e^p,e^\bfx),
\ee
where $\bfmu\in \frac{1}{2}c_1 + H^2(M,\BZ)$ equals half the Stiefel-Whitney class of the bundles. 

The work of G\"ottsche \cite{Gottsche:1996} and
G\"ottsche-Zagier \cite{Gottsche:1996aoa} connected Donaldson invariants to the subject of modular forms,
which is at first sight rather distant from the above. G\"ottsche and
Zagier realized that generating functions of Donaldson invariants of
rational surfaces, as determined earlier for example in \cite{LiQin1993, Kotschick1995, ellingsrud1995wall},
could be written as a residue of a combination of modular
forms and so-called indefinite theta functions. The latter enjoy much scientific interest in
recent years, in part due to their connection to Ramanujan's mock theta
functions \cite{ZwegersThesis, MR2605321}. An indefinite theta function $\Theta: \BH\to \BC$ is a holomorphic $q$-series defined as a sum over an
indefinite lattice $\Lambda$ with signature $(1,n-1)$. The sum is convergent, since the sum is
restricted to lattice points with negative definite norm (for the convention
taken in this paper). However, $\Theta$ does not transform as a
modular form under $\operatorname{SL}(2,\BZ)$ transformations. The latter
can however be cured thanks to the seminal work of Zwegers': one may
add a specific non-holomorphic function $R$ to $\Theta$ such that 
the sum $\widehat \Theta=\Theta +R$ transforms as a modular form. Interestingly, 
the $\bar \tau$-derivative $\Psi=\partial_{\bar \tau} \widehat
\Theta$, turns out to be a Siegel-Narain theta function associated to $\Lambda$, whose modular
properties are more easily determined using the familiar Poisson resummation.  

Let us now return to the physical $u$-plane integral to explain the
main result of this paper. The integral can be expressed as an integral over the fundamental domain $\BH/\Gamma^0(4)$, where $\Gamma^0(4)\in
\operatorname{SL}(2,\BZ)$ is the electric-magnetic duality group, after
a change of variables from $u$ to the
effective coupling constant $\tau$. Subsequently the technique of
``lattice reduction'' can be applied to evaluate the integral when $b_2(M)>1$ \cite{Moore:1997pc, LoNeSha}. This technique was originally developed in
the context of one loop amplitudes in string theory
\cite{Dixon:1990pc, Harvey:1995fq} and also has major mathematical
applications \cite{Borcherds:1996uda}. For the manifold
$\mathbb{P}^2$, with $b_2(\mathbb{P}^2)=1$, the integrand was realized
as a total derivative to $\bar \tau$ using Zagier's modular completion
of the class number generating function \cite{zagier:1975}. 

The $u$-plane integral provided in this way a physical
derivation of the results on Donaldson invariants by G\"ottsche and
Zagier \cite{Gottsche:1996, Gottsche:1996aoa}, and reproduced in
particular generating functions for rational
surfaces including $\mathbb{P}^1\times \mathbb{P}^1$, the
wall-crossing formula and the blow-up formula \cite{Moore:1997pc}. Later work by Griffin, Malmendier and Ono
proved agreement \cite{Malmendier:2008db,
  Malmendier:2010ss, Griffin:2012kw} in additional cases including $\mathbb{P}^2$. 
  
The present paper demonstrates that the $u$-plane integral can be
evaluated quite generally by expressing it as a total derivative of an
indefinite theta function. To this end, we add a
$\mathcal{\bar  Q}$-exact term to the effective action of the
Donaldson-Witten theory used in \cite{Moore:1997pc}, which does not
modify the value of the integral by the usual rules of topological
field theory. Using techniques of indefinite 
theta functions developed by Zwegers \cite{ZwegersThesis}, we show that the modified integrand is a
total $\bar \tau$-derivative for an arbitrary four-manifold with $b_2^+=1$. The
integrand of the $u$-plane integral equals what is known as the ``shadow'' of the
indefinite theta series (up to an overall multiplicative function). As a
result, the $u$-plane integral can be immediately evaluated for a generic choice of
metric, and reproduces precisely G\"ottsche's results for complex algebraic manifolds. The same
technique can be applied when matter is included, and we hope similar  
techniques can be developed for gauge groups with rank larger than
one, theories of class $\mathcal{S}$ \cite{Gaiotto:2009we, Gaiotto:2009hg} or more general non-Lagrangian theories. We moreover expect
that it may be applied more widely as an alternative for ``lattice reduction'' to evaluate
modular integrals.

From a more general point of view, it is interesting to note that the Coulomb branch integral 
provides both the holomorphic and non-holomorphic terms of the indefinite theta function. In other cases where
such mock modular forms appear in physics, such as in Vafa-Witten theory \cite{Vafa:1994tf}, AdS$_3$ gravity
\cite{Manschot:2007ha}, black holes \cite{Manschot:2009ia,
  Dabholkar:2012nd, Alexandrov:2016tnf}, or the moonshine phenomenon
\cite{Cheng:2011ay}, the holomorphic part has usually the clearest
physical interpretation, whereas the non-holomorphic term is
typically less well understood.

We conclude the introduction with an outline of the article. For a
self-contained presentation, Section \ref{SectionSW}
reviews relevant aspects of Seiberg-Witten theory, and Section
\ref{sec_uplane} reviews the
path integral of the topologically twisted theory. Section
\ref{sec_uplane} also discusses the $\bar{\CQ}$-exact surface
operator, and how it modifies the integrand of the $u$-plane integral.
In Section \ref{sec_integration}, we evaluate the integral and provide
explicit results for a few complex rational surfaces. In Section \ref{higherrank} we discuss the generalization of
$\bar{\CQ}$-exact surface operator for higher rank gauge groups and
how it modifies the integrand of the Coulomb branch integral. We
conclude with a brief discussion and concluding remarks. The Appendices contain
relevant properties of modular forms and (indefinite) theta functions.

\section{Seiberg-Witten geometry} \label{SectionSW}
We begin our discussion with a brief reminder of the solution of
Seiberg and Witten of the SU(2) $\CN=2$ super Yang-Mills gauge theory
on $\BR^4$ \cite{Seiberg:1994rs, Seiberg:1994aj} (see \cite{AlvarezGaume,
  Bilal:1995hc, Shnir,Tern09} for a review, \cite{Tachikawa13,Gaiotto:2014bja} for a more modern perspective, and
\cite{Nakajima:2003uh,Bouwknegt02} for a more mathematically inclined
discussion).

\subsection{The Seiberg-Witten solution} 

The $\CN=2$  vector multiplet contains a scalar field $\phi$,
two Weyl fermions, $\lambda$, $\psi$, and the gauge connection $A$,
which are all valued in the adjoint representation of SU(2). For
supersymmetry to be closed off-shell we include a real auxiliary scalar $V$ and a complex auxiliary scalar $W=W_1+iW_2$ \cite{Tern09}. The potential of the theory is $V(\phi)=g^{-2}\,\text{Tr}[\phi,\phi^{\dagger}]^2$ and this is minimized by $\phi = a\sigma_3$ up to a gauge transformation. The Weyl group of SU(2) acts by $a \to -a$, thus $u=\frac{1}{16\pi^2} \mathrm{Tr}\, \phi^2$ is an invariant quantity and functions as a coordinate on the Coulomb branch moduli space. Quantum-mechanically, we let $u$ be the expectation value
$$u(a)=\frac{1}{16\pi^2} \left< \mathrm{Tr}\, \phi^2\right>.$$
The Coulomb branch of the quantum moduli space $\CM_{\rm C} $ is isomorphic to $\BP^1 \backslash \{\infty, \pm \Lambda^2 \}$ with $\Lambda$ being a dynamically generated scale. At the points $u=\pm \Lambda^2$, the effective Coulomb branch theory becomes singular since either the monopole or dyon, which were integrated out in the effective theory, becomes massless. 

The BPS states of the theory carry both electric charge $n_{\rm e}$ and magnetic charge $n_{\rm m}$ that belong to a charge lattice, $(n_{\rm e}, n_{\rm m})  \in \Gamma_u \cong \BZ \times \BZ$. The complexified mass (or central charge) for a dyonic state is
$$
Z_u=n_{\rm e} a+ n_{\rm m} a_D.
$$
At weak coupling $g^2\to 0$, these charges are related to the Coulomb branch parameter $u$ and the effective complexified gauge coupling $\tau=\frac{\theta}{\pi}+\frac{8\pi\,i}{g^2} \in \BH$ by the relations
$$
a(u)=\sqrt{\frac{u}{2}},\qquad \qquad  a_D(u)=\tau a(u).
$$

In the seminal work by Seiberg and Witten \cite{Seiberg:1994rs,
  Seiberg:1994aj}, $u$, $a$ and $a_D$ are determined for arbitrary
coupling constant $\tau$, making use of an elliptic curve $\Sigma$,
whose complex structure is identified with $\tau$. The Coulomb branch
corresponds to a family of elliptic curves, which are identified as the solution sets to
the following algebraic equation \cite{Seiberg:1994aj}
\be
\label{SWcurve}
y^2=4x(x^2-ux+\tfrac{1}{4}\Lambda^4), \qquad x,y\in \BC,
\ee
where $\Lambda$ is the dynamically generated mass scale of the theory. One can determine $u$ in terms of $\tau$ using techniques from the theory of elliptic curves. Substituting for $x=\alpha^{-2} \tilde x+\frac{1}{3}u$ and $y=\alpha^{-3}\tilde y$ brings the curve in Weierstrass form 
\[
\tilde y^2=4\tilde x^3-g_2 \tilde x -g_3.
\]
The constants $g_2$ and $g_3$ are given in terms of Eisenstein series $E_k$
as $g_2=\frac{4 \pi^4}{3} E_4$ and $g_3=\frac{8 \pi^6}{27} E_6$. See
Equation (\ref{Ek}) in Appendix \ref{app_mod_forms} for the definition
of the $E_k$. The parameter $\alpha$ can be expressed as  
$$
\alpha=\frac{\sqrt{2}\pi}{\Lambda} \vartheta_2\vartheta_3,
$$ 
where $\vartheta_j$ are the Jacobi theta
functions defined in (\ref{Jacobitheta}) in Appendix \ref{app_mod_forms}.\footnote{Our conventions are such that for $\Lambda=1$, $u$
  has the same $q$-expansion as in \cite{Moore:1997pc,  LoNeSha, Seiberg:1994aj}.} This gives for $u=u(\tau)$ in terms of Jacobi theta functions
\be
\label{modu}
\frac{u}{\Lambda^2}=\frac{\vartheta_2^4+\vartheta_3^4}{2\vartheta_2^2\,\vartheta_3^2}=\tfrac{1}{8}q^{-\frac{1}{4}}+\tfrac{5}{2}
q^\frac{1}{4}-\tfrac{31}{4}q^{\frac{3}{4}}+O(q^{\frac{5}{4}}),
\ee
where $q=e^{2\pi i \tau}$. The function $u$ is left invariant by a
congruence subgroup of the full modular group SL$(2,\BZ)$. Using the
transformation properties of the Jacobi theta functions (\ref{Jacobitheta_trafos}), one
verifies that the congruence subgroup is $\Gamma^0(4)$ of SL$(2,\BZ)$ (\ref{Gamma04}).

The discriminant $\Delta$ of the curve is proportional to
$u^2-\Lambda^4$, and vanishes or diverges when the curve
(\ref{SWcurve}) is singular giving three distinct \emph{degenerations}
of the elliptic curve. Physically, a monopole or dyon becomes
massless for $\Delta=0$. These singular points should be excluded from
the Coulomb branch, and this is the reason which lead us to the identification of $\CM_{\rm C}$ with
$\BP^1 \backslash \{ \pm \Lambda^2, \infty \}$ as explained earlier. We may also
parametrize the Coulomb branch as a coset of the upperhalf plane,
since $u$ is a modular form of $\Gamma^0(4)$. Therefore,
$\CM_{\rm C}$ can alternatively be identified with $\BH /
\Gamma^0(4)$. This fundamental domain has three cusps for $\tau\to i
\infty$, $\tau\to 0$ and $\tau\to 2$, which correspond respectively to
weak coupling, $u=\Lambda^2$ and $u=-\Lambda^2$. 

Using rigid special geometry, $a_D$ is derived from the prepotential $\CF$  that specifies the low energy effective action of the theory,
$$
\CF(a,\Lambda)=\frac{4i}{\pi} a^2 \log\!\left(\frac{a}{\Lambda}\right) + a^2 \sum_{k=0}^\infty c_k \left(\frac{\Lambda}{a}\right)^{4k}.
$$ 
The second summand corresponds to instanton corrections. The magnetic
dual of $a$ is expressed in terms of $\CF$ as
$$
a_D=\frac{\partial \CF}{\partial a},
$$
and for the complexified gauge coupling we have
$$
\tau=\frac{\partial a_D}{\partial a}.
$$
Seiberg and Witten expressed $a$ and $a_D$ as integrals of a differential $\lambda$ over one-cycles of the curve $\Sigma$. To this end, let $A$ and $B$ form a symplectic basis of $H_1(\Sigma, \BZ)$. Then $a$ and $a_D$ are given as
$$
a=\int_A \lambda,\qquad \qquad a_D=\int_B \lambda.
$$
The right behavior at the singularities and positivity of the metric imply that the differential $\lambda$ satisfies
$$
\frac{d\lambda}{du}=\frac{\sqrt{2}}{4\pi}\frac{dx}{y} \in H^1(\Sigma, \BC).
$$
Choosing a complex coordinate on the curve $\Sigma$, $z\in \BC/\{m\tau+n\}$, $m,n\in \BZ$, we let the $A$-cycle correspond to the straight line connecting $0$ to $1$, and the $B$-cycle the straight line from $0$ to $\tau$.  To express $a$ in terms of modular forms, we use the map of the parametrisation (\ref{SWcurve}) of the Seiberg-Witten curve to the Weierstrass form. Then $(\tilde x,\tilde y)=(\wp(z),\wp'(z))$. As a result one finds that 
\be
\label{dau}
\frac{da}{du}=\frac{1}{2\Lambda}\vartheta_2\vartheta_3.
\ee
Integrating with respect to $u$ expresses $a$ in terms of
(quasi)-modular forms
\be 
\label{eq:a}
\frac{a}{\Lambda}=\frac{2E_2+\vartheta_2^4+\vartheta_3^4}{6\,\vartheta_2\,
  \vartheta_3}=\tfrac{1}{4}q^{-\frac{1}{8}}+\tfrac{3}{2}q^\frac{3}{8}-\tfrac{21}{4}q^{\frac{7}{8}}+O(q^{\frac{11}{8}}).
\ee
See Equation (\ref{Ek}) in Appendix \ref{app_mod_forms} for the definition of
$E_2$.

\subsection{The topologically twisted theory}
\label{sec_top_twist}
Topological twisting allows to arrive at a topological quantum field
theory starting from a theory with extended supersymmetry \cite{Witten:1988ze}.
After topologically twisting the holonomy group with the SU(2)$_R$
R-symmetry, one of the original supersymmetry generators transforms
trivially under the twisted holonomy group of the four-manifold. This
generator, denoted by $\mathcal{\bar Q}$, is the BRST operator of the
theory and all observables are $\mathcal{\bar Q}$-closed. Topological
twisting changes the representations of the field content under the
rotation group. The gauge connection $A$ and complex scalar $a$ remain
respectively a 1-form and a scalar, however the fermionic fields are
now a 0-form $\eta$, 1-form $\psi$ and self-dual 2-form $\chi$. The
three real auxiliary fields $V$, $W_1$ and $W_2$ combine to an auxiliary self-dual two-form $D$
\cite{Moore:1997pc}. The action of $\mathcal{\bar Q}$ on these fields is given by
\be
\label{barQcomm}
\begin{split}
&[\mathcal{\bar Q},A]=\psi, \hspace{55pt} [\mathcal{\bar Q},a]=0, \hspace{50pt} [\mathcal{\bar Q},\bar a]=\sqrt{2}i \eta,  \\
&   [\mathcal{\bar Q},D]=(d_A\psi)_+, \qquad  \{\mathcal{\bar Q},\psi\}=4\sqrt{2}\,da, \\
& \{\mathcal{\bar Q},\eta\}=0, \hspace{54pt}    \{\mathcal{\bar Q},\chi\}=i(F_+-D),
\end{split}
\ee 
where the subscript $+$ indicates the self-dual component of two-form,
thus $F_+=\frac{1}{2}(F+*F)$. The four-manifolds considered in this
paper have $b_2^+(M)=1$. The self-dual part of the curvature
$[F_+]\in H^2(M)$ is for such a four-manifold proportional to the self-dual harmonic form $J\in H^2(M)$. 

For completeness we present the full Lagrangian $\mathcal{L}$ of the twisted theory on $\mathbb{R}^4$ which is given by \cite{Moore:1997pc}
\be
\label{Lagrangian}
\begin{split}
\mathcal{L}=&\frac{i}{16\pi}(\bar \tau F_+\wedge F_++\tau F_-\wedge F_- )+\frac{\tau_2}{8\pi}\,da\wedge *d\bar a-\frac{\tau_2}{8\pi} D\wedge *D\\
&-\frac{1}{16\pi}\tau \psi \wedge *d\eta+\frac{1}{16\pi}\bar \tau \eta\wedge d*\psi+\frac{1}{8\pi}\tau\psi\wedge d\chi-\frac{1}{8\pi}\bar \tau \chi\wedge d\psi\\
&+\frac{i\sqrt{2}}{16\pi}\frac{d\bar \tau}{d\bar a}\,\eta\chi \wedge (F_++D)-\frac{i\sqrt{2}}{2^7\pi} \frac{d\tau}{da}\psi\wedge \psi\wedge (F_-+D)\\
&+\frac{i}{3\cdot 2^{11}} \frac{d^2\tau}{da^2}\psi\wedge\psi\wedge\psi\wedge\psi -\frac{\sqrt{2}i}{3\cdot 2^5\pi}\left\{\mathcal{\bar Q},\chi_{\mu\nu}\chi^{\nu\lambda}\chi_{\lambda}^{\,\,\, \mu} \right\}\sqrt{g}\,d^4x.
\end{split}
\ee
On curved manifolds, the theory contains further couplings to the
background curvature, which were first derived by \cite{Witten:1995gf}
based on the R-symmetry anomaly. On toric four-manifolds, one may
derive these terms from gravitational couplings in the Nekrasov
partition function \cite{Ne, Nakajima:2003uh}. As we are working in a
topologically twisted theory, the terms can only involve the Euler
characteristic $\chi(M)$ and signature $\sigma(M)$ of $M$. 
For the four-manifolds considered in this paper, $\chi(M)+\sigma(M)=4$. These contributions can be gathered in the \emph{measure factor} which takes the form
\begin{equation}
\label{nutau0}
\nu(\tau)=8i(u^2-1)\frac{d\tau}{du}\left(  \frac{( \frac{2i}{\pi} \frac{du}{d\tau} )^2}{u^2-1} \right)^{\sigma(M)/8},
\end{equation}
where the multiplicative constants are chosen to match the conventions for  
Donaldson invariants. Using the identity \cite{Matone:1995rx}
\begin{equation} \label{u2minus1}
u^2-1=\frac{i}{4\pi }  \frac{du}{d\tau} \left( \frac{du}{da} \right)^2,
\end{equation}
equation (\ref{nutau0}) equals to
\be
\label{nutau}
\nu(\tau)=-\frac{2^{\frac{3\sigma(M)}{4}+1}}{\pi} (u^2-1)^{\frac{\sigma(M)}{8}}\left(\frac{da}{du} \right)^{\frac{\sigma(M)}{2}-2}.
\ee

The observables of the topological theory lie in the $\mathcal{\bar Q}-$cohomology. The $k-$form observables $\mathcal{O}^{(k)}$ relevant for Donaldson theory are obtained as a solution to a descent prescription \cite{Moore:1997pc,Witten90},  
$$ 
d\CO^{(k)}=\{\mathcal{\bar Q},\CO^{(k+1)}\}.
$$
The operator obtained by integrating $\CO^{(k)}$ over a $k$-cycle is then automatically $\mathcal{\bar Q}$-closed. The descent equations can be solved using an operator $K$, satisfying $\{\mathcal{\bar Q}, K\}=d$, such that $\CO^{(k)}=K^k\CO^{(0)}$. We choose $\CO^{(0)}=\frac{1}{8 \pi^2} \text{Tr}\,\phi(P)^2$ with $P$ a point of $M$. In the effective, topological theory $\phi$ is constant over $M$, and therefore $e^{p\CO^{(0)}(P)}=e^{2pu}$. Using the descent procedure, one finds for the surface observable $\CO^{(2)}$ \cite{Moore:1997pc, LoNeSha}
$$
I(\bfx)=\frac{1}{4\pi^2} \int_\bfx \mathrm{Tr}\!\left[\frac{1}{8} \psi\wedge \psi -\frac{1}{\sqrt{2}}\phi F \right],
$$
with $\bfx\in H^2(M)$. In the low energy effective theory, $\CO^{(0)}$ is given by the Seiberg-Witten solution, $\CO^{(0)}=u$, and the surface operator is modified to \cite{Moore:1997pc}
\be
\label{I-}
\widetilde I_-(\bfx)=\frac{i}{ \sqrt{2}\pi}\int_{\bfx} \frac{1}{32}\,\frac{d^2u}{da^2}\,\psi\wedge \psi-\frac{\sqrt{2}}{4}\frac{du}{da}(F_-+D),
\ee
\noindent where $F$ is field strength of the remaining $U(1)$ gauge
symmetry. To evaluate the $u$-plane integral using indefinite theta
functions, we will add to this surface operator a $\mathcal{\bar
  Q}$-exact operator 
\be
\label{I+0}
\widetilde I_+(\bfx)=-\frac{1}{4\pi}\int_{\bfx}\left\{\mathcal{\bar Q}, \frac{d\bar u}{d\bar a}\,\chi\right\}.
\ee
Using (\ref{barQcomm}), this evaluates to 
\be
\label{I+}
\widetilde I_+(\bfx)= -\frac{i}{ \sqrt{2}\pi}  \int_{\bfx}  \frac{1}{2} \frac{d^2\bar u}{d \bar a^2}\,\eta \,\chi+ \frac{\sqrt{2}}{4} \frac{d\bar u}{d \bar a}(F_+-D).
\ee
This term couples to the self-dual part $F_+$ of $F$, whereas
(\ref{I-}) involved only $F_-$. Further considerations of such
modifications of the topological action are
discussed in \cite{KMMN}. 
    
Finally, the renormalization group flow to the low energy theory gives
rise to a contact term \cite{Moore:1997pc, LoNeSha}, $G\,\bfx^2$, which is a consequence of
the self-intersection of the cycle $\bfx$ appearing in the surface operators. Since
the surface operators $\widetilde I_{\pm}$ are $\mathcal{\bar
  Q}$-closed, the coefficient $G$ of the contact term is necessarily
holomorphic in $u$. It is conveniently expressed in terms of (quasi)-modular forms
\begin{equation} \label{ContactTerm}
\begin{split}
G(u)&=\frac{1}{24}\left( 8 u -E_2 \left(\frac{du}{da}
  \right)^2\right),
\end{split}
\end{equation}
where $E_2$ is the Eisenstein series of weight 2 defined by Eq. (\ref{Ek}) in the Appendix, or as a derivative to $\vartheta_4$ \cite{LoNeSha}
$$
G(u)=-\frac{1}{2\pi i} \left( \frac{du}{da}\right)^2 \partial_{\tau} \log \vartheta_4 .
$$
If we want to emphasize the dependence of $G$ on $\tau$, we will write sometimes $G(\tau)$ instead of $G(u)$.


\section{The $u$-plane integral} \label{sec_uplane} 
The $u$-plane integral is the path integral over the Coulomb branch of  topologically twisted
$\CN=2$ supersymmetric gauge theory with gauge group SU(2) or 
SO(3). In this section we will explicitly describe the $u$-plane 
integral following mostly \cite{Moore:1997pc}. See for an overview
also \cite{Labastida:2005zz}. Let us start with a few comments on the four-manifold
$M$, which we choose to be a simply connected ($b_1=0$) manifold,
compact without boundary. Integration on $M$ gives $H^2(M,\BZ)$
naturally the structure of a lattice $\Lambda\cong\BZ^{b_2}$ with unimodular quadratic form $Q : H^2(M) \to \BZ$ and bilinear form $B: H^2(M) \times H^2(M) \to \BZ$ 
\be
\label{qform_notation}
Q(\bfk)\equiv\bfk^2 \equiv \int_M \bfk \wedge \bfk, \qquad 
B(\bfk_1,\bfk_2)\equiv \int_M\bfk_1\wedge \bfk_2,
\ee
which has signature $(b_2^+,b_2^-)$. We project $\bfk$ to the positive
definite and negative definite subspace as explained in Appendix
\ref{app_mod_forms}. Viewed as a 2-form, $\bfk_+$ is self-dual under the Hodge
$*$-operation, while $\bfk_-$ is anti-self-dual.  

We restrict in the following to four-manifolds with $b_2^+=1$. The corresponding lattices $\Lambda$ are completely
classifield. If $\Lambda$ is odd, the matrix associated to $Q$ can be brought to
the diagonal form
\be  
\label{Qodd}
\left<1\right>+m\left<-1\right>,
\ee
with $m=b_2-1$. If $\Lambda$ is even, the matrix associated to $Q$ is equivalent to
\be
\label{Qeven}
\left(\begin{array}{cc} 0 & 1 \\ 1 & 0 \end{array}\right) \oplus n \Lambda_{E_8},
\ee
where $\Lambda_{E_8}$ denotes the root lattice of the $E_8$ group, and $n=(b_2-2)/8$.

\subsection{The topologically twisted path integral}\label{uplanemodforms} 
The path integral of the effective theory on the Coulomb branch is
given by 
\be  
\label{path_int}
\Phi_{\bfmu}^J(p,\bfx)= \int [\mathcal D X]\,\nu(\tau)\,e^{-\int_{M}\mathcal{L}+2pu+\tilde I_-(\bfx)+\tilde I_+(\bfx)+\bfx^2\,G},
\ee
where $[\mathcal D X]$ stands for the path integral measure of the
fields $[\cD(A,a,\bar a,\eta, \chi,\psi,D)]$. As mentioned in the
introduction and will be confirmed
in the following, $\Phi_{\bfmu}^J$
depends discontinuously on the metric $g$, and may jump across walls of marginal
stability due to the presence of Abelian instantons. The metric
dependence of $\Phi_{\bfmu}^J$ is only through the period point $J=J(g)$
\cite{Kotschick1995}. 

To evaluate the path integral, we start by integrating out $D$. We do
so by substituting for $D$ the solution to its equation of
motion. From the terms in the action (\ref{Lagrangian}) involving $D$,
together with the $D$-dependent terms in (\ref{I-}) and (\ref{I+}),
one finds 
\be
\label{identityD}
D=-\frac{2\,\mathrm{Im}\!\left(du/da\right)}{\tau_2}\, \bfx_++\frac{\sqrt{2}i}{ 4 \tau_2}\frac{d\bar \tau}{d\bar a}\,\eta\chi,
\ee
where $\bfx$ denotes the two-form Poincar\'e dual to the cycle $\bfx$
in (\ref{I-}) and (\ref{I+}), or equivalently $\int_{\bfx} \omega = \int_M
\bfx\wedge \omega $ for any 2-form $\omega$. To simplify notation, we
define the variables $\bfrho\in H^2(M,\mathbb{C})$ and $\bfb\in H^2(M,\mathbb{R})$ by 
\be
\label{rho}
\bfrho=\frac{\bfx}{2\pi}\frac{du}{da},\qquad \qquad \bfb=\frac{\mathrm{Im}(\bfrho)}{\tau_2}.
\ee
After substitution of (\ref{identityD}) back in the $D$-dependent
terms of the Lagrangian and surface operators, they contribute 
\be 
\label{expD2}
\exp\!\left(-2\pi \tau_2 \bfb_+^2 +\frac{i\sqrt{2}}{4} \frac{d\bar \tau}{d\bar a}\, \int_M \bfb_+\wedge\eta\,\chi\right)
\ee  
to the integrand of the path integral.

Next we integrate over the fermionic fields. To understand the
contribution to the $u$-plane integral from these fields it is useful
to discuss the scaling of the fields under a rescaling of the metric: $\lim_{t\to
  \infty} = t^2g_0$ for a fixed metric $g_0$.\footnote{This 
  one-parameter family of metrics belongs to a single chamber.} The scaling dimensions of the zero modes naturally equals their form
degree.  These equal the scaling dimensions of the quantum
fluctuations of the fields, except for $\eta$, whose quantum
fluctuation has dimension 2 instead of 0 \cite[Section
2.3]{Moore:1997pc}.  Thus we see that the terms of the Lagrangian involving $\eta$ and
$\chi$ have scaling dimension larger than four, except when we replace
$\eta$ by its zero-mode $\eta_0$. Similarly, the term involving $\eta$
in the surface operator (\ref{I+}) has dimension 2 if we replace $\eta$
by its zero mode. One can show that the corrections due to the quantum fluctuations of
$\chi$ do not survive in the limit $t\to\infty$, assuming that
$b_1(M)=0$.  

Therefore, the path integral over the fermionic fields is reduced to the integral over zero modes.
Collecting the terms involving the zero modes gives
\be
\begin{split}
&\int [d\eta_0\,d\chi_0]\,\exp\!\left( -\frac{\sqrt{2}i}{16\pi}\frac{d\bar \tau}{d\bar a}\int_M\eta_0\,\chi_0 \wedge (F_+-4\pi \bfb_+)  -\frac{i}{\pi \sqrt{2}}  \int_{\bfx}  \frac{1}{2} \frac{d^2\bar u}{d \bar a^2}\,\eta_0 \,\chi_0 \right), 
\end{split}
\ee
Carrying out the integral gives
\be
\int_M \left(\frac{\sqrt{2}i}{16\pi} \frac{d\bar \tau}{d\bar a}\left(F
    -4\pi \bfb\right)
  +\frac{i}{\sqrt{2}} \frac{d\bar \bfrho}{d\bar a}\right)\wedge
\underline J,
\ee
where $\underline J$ is the normalization of $J$, $J/\sqrt{Q(J)}$. This can interestingly be written in a simpler form as a derivative to $\bar \tau$,
\be
\label{etachi_term}
\frac{\sqrt{\tau_2}}{4\pi}\,  \frac{d\bar \tau}{d\bar
  a}\,\partial_{\bar \tau}\sqrt{2\tau_2}\, B(F +4\pi \bfb, \underline J),
\ee
where we used the notation (\ref{qform_notation}), and $\partial_{\bar
\tau}$ acts on all terms to its right.

Finally, the photon path integral contains a sum over all fluxes
times a factor of
$\tau_2^{-\frac{1}{2}}$\cite{Witten:1995gf}.\footnote{Here we assumed
  $b_1=0$; for a non-simply connected four-manifold this factor
  equals $\tau_2 ^{(b_1-1)/2}$.} The U(1) fluxes $[F]/4\pi$ lie
in a shift of the integer cohomology group $H^2(M,\mathbb{Z})$ by 
half the second Stiefel-Whitney class $w_2(E)$ of the $\operatorname{SU}(2)$ or SO(3)  bundle $E$, since $F$ is the field strength of the
unbroken U(1) gauge group. We introduce the conjugacy class $\bfmu\in
H^2(M,\mathbb{Z}/2\BZ)$, such that $w_2(E)=2\bfmu+H^2(M,2\mathbb{Z})$ and $\frac{1}{4\pi}[F]\in H^2(M,\BZ)+\bfmu$.
With $\bfk=\frac{1}{4\pi}[F]$, the photon path integral can be written concisely in terms of a Siegel-Narain theta series $\Psi_\bfmu^J(\tau,\bfrho)$ defined as
\be
\label{Psi}
\begin{split}
\Psi^J_{\bfmu}(\tau, \bfrho
)&=\exp\!\left(-2\pi\tau_2 \bfb_+^2\right)\sum_{\bfk
  \in \Lambda +\bfmu} \partial_{\bar
  \tau}\left(\sqrt{2\tau_2}B(\bfk+\bfb, \underline J)\right) \\ 
                                         & (-1)^{B(\bfk, K_M)} \,\exp\!\left( -i\pi \bar \tau \bfk_+^2 - i\pi \tau \bfk_-^2 -2\pi i  B(\bfk_+, \bar \bfrho) - 2\pi i B(\bfk_-,\bfrho) \right), 
\end{split}
\ee  
where we identify $\Lambda$ with $H^2(M,\BZ)$.
The first exponential on the right hand side of (\ref{Psi}) is due to
(\ref{expD2}), and the first term after the summation sign due to
(\ref{etachi_term}) divided by $\sqrt{\tau_2}$. We recognize the couplings in the exponent on the second line as due to the classical Yang-Mills action \cite{Witten:1995gf, Verlinde:1995mz} and the surface operators $\tilde I_+$ and $\tilde I_-$. The sign $(-1)^{B(\bfk, K_M)}$, with $K_M$ the canonical class of $M$, arises from integrating out the massive fermions in the Coulomb branch of the topologically twisted theory \cite{Witten:1995gf}. 

The path integral with the insertions of the point and surface operator then equals
\be
\Phi_{\bfmu}^J(p,\bfx) =\int_{\CM_C} da\wedge d\bar a\,\nu(\tau)\,\frac{d\bar \tau}{d\bar a} \, \Psi_\bfmu^J(\tau,\bfrho)\,e^{2pu +\bfx^2\,G}.
\ee 
The integration domain is most naturally stated in terms of $\tau$,
rather then $a$. Since the duality group of Seiberg-Witten theory is
$\Gamma^0(4)$, the domain for $\tau$ is naturally
$\BH/\Gamma^0(4)$. We then arrive at the following modular integral
\be
\label{Phi_final}
\Phi^J_{\bfmu}(p,\bfx)=\int_{\BH/\Gamma^0(4)} d\tau\wedge d\bar \tau
\,\tilde \nu(\tau) \, \Psi_\bfmu^J(\tau,\bfrho)\,e^{2pu +\bfx^2\,G},
\ee 
where we defined
\be
\tilde \nu=\nu\,\frac{da}{d\tau}. 
\ee

\subsection{Modular invariance of the integrand}
For completeness, we discuss in this subsection invariance of the
integrand under the $\Gamma^0(4)$ duality group of Seiberg-Witten
theory, which is an important consistency requirement for the integrand. Since
$d\tau \wedge d\bar \tau$ transforms with weight $(-2,-2)$, the
integrand in (\ref{Phi_final}) must have modular weight $(2,2)$. Let us start with the function $\Psi_\bfmu^J$
(\ref{Psi}) of the integrand. This is an example of a Siegel-Narain
theta function. A general form of such theta functions which suits our
purposes is given in Appendix 
\ref{app_mod_forms}, Equation (\ref{PsiJ}). To compare
(\ref{Phi_final}) with that equation, we set $\bfz=\bfrho$ and $\bfb=\mathrm{Im}(\bfrho)/\tau_2$, with
$\bfrho$ as defined in (\ref{rho}). Furthermore, we identify $K$
in (\ref{PsiJ}) with the canonical class $K_M$, which is a characteristic
element of $H^2(M,\BZ)$.\footnote{This follows for example from the Hirzebruch-Riemann-Roch theorem for a line bundle.}
The transformation properties of $\Psi^J_\bfmu$ under
$\operatorname{SL}(2,\BZ)$ and $\Gamma^0(4)$ are given in
(\ref{Psi_trafos}) -- (\ref{PsiT4}). 

We see that $\bfrho$ appears in $\Psi_\bfmu^J$ as an elliptic variable. Indeed, since
$\frac{da}{du}$ from equation (\ref{dau}) is a modular form of weight $1$ under
$\Gamma^0(4)$, $\bfrho$ transforms as an elliptic variable. More
precisely, one verifies using the properties of the Jacobi
theta functions (\ref{Jacobitheta_trafos}), that $\bfrho$ transforms
under the two generators of $\Gamma^0(4)$ as
\be 
\label{rhotrafo}
\begin{split} 
&\bfrho(\tau+4)=-\bfrho(\tau), \qquad \qquad \bfrho\!\left(\frac{\tau}{\tau+1} \right)=\frac{\bfrho(\tau)}{\tau+1}.
\end{split}
\ee 
Note that this differs by a minus sign from the usual transformation
of an elliptic variable under $\tau\to \tau+4$. 

With these transformations, we can determine the action of $\Gamma^0(4)$ generators on
$\Psi_\bfmu^J(\tau,\bfrho)$. Recall
that $\bfmu\in H^2(M,\BZ/2)$ in the path integral. Combining
(\ref{Psi-1}) and (\ref{PsiT4}), we find for the generator $\tau\to \tau+4$ of $\Gamma^0(4)$ 
\be
\label{psitau4}
\left.\Psi^J_{\bfmu}(\tau,\bfrho)\right|_{\tau\mapsto \tau+4}=- \Psi^J_{\bfmu}(\tau, \bfrho).
\ee
Using (\ref{PsiS}), we derive for the action of the second generator
\be
\label{psitautau1}
\left.\Psi^J_{\bfmu}\!\left( \tau,\bfrho\right)\right|_{\tau\mapsto \frac{\tau}{\tau+1}}=(\bar \tau+1)^2(\tau+1)^{\frac{b_2}{2}}\exp\!\left(-\frac{\pi i\bfrho^2}{\tau+1} +\frac{\pi i}{4}\sigma(M)\right) \Psi^J_{\bfmu}(\tau,\bfrho),
\ee
where we used $K_M^2=8+\sigma(M)$ for simply connected four-manifolds
with $b_2^+=1$.  

Next we discuss the contact term $e^{\bfx^2 G}$ with
$G$ given in (\ref{ContactTerm}). Due to the special
transformations of the weight two Eisenstein series $E_2$ given in (\ref{E2trafo}), the contact term transforms
as follows
\be
\label{contact_trafo}
e^{\bfx^2 G(\tau+4)}=e^{\bfx^2 G(\tau)},\qquad e^{\bfx^2
  G\left(\frac{\tau}{\tau+1}\right)}=e^{\bfx^2 G(\tau)+\frac{\pi
    i}{\tau+1}\bfrho^2}. 
\ee 

The remaining term in the integrand is $\tilde \nu$. Using the identity 
\be
\left(\frac{\left(\frac{2i}{\pi} \frac{du}{d\tau}\right)^2}{u^2-1}\right)^{\frac{1}{8}}=\vartheta_4(\tau),
\ee 
we can write $\tilde \nu$ as 
\be
\label{tildenu}
\tilde \nu(\tau) = -8i(u^2-1)
\frac{da}{du}\,\vartheta_4(\tau)^{\sigma(M)}. 
\ee
If we express further $u$ and $da/du$ in terms of Jacobi theta
functions and use the transformation properties (\ref{Jacobitheta_trafos})
under $\Gamma^0(4)$, one finds 
\be
\label{nu_trafo}
\tilde \nu(\tau+4)=-\tilde \nu(\tau), \qquad
\tilde
\nu\!\left(\frac{\tau}{\tau+1}\right)=(\tau+1)^{2-\frac{b_2(M)}{2}}
e^{-\frac{\pi i \sigma(M)}{4}}\tilde \nu(\tau).
\ee
Combining now (\ref{psitau4}), (\ref{psitautau1}),
(\ref{contact_trafo}) and (\ref{nu_trafo}), we conclude that the integrand of (\ref{Phi_final}) has indeed weight (2,2) under $\Gamma^0(4)$
as required.


\section{Evaluation of the $u$-plane integral} \label{sec_integration}

The previous section reduced the path integral (\ref{path_int}) to the
integral (\ref{Phi_final}) over the fundamental domain $\BH/\Gamma^0(4)$. This domain is the
union of six images of the SL$(2,\BZ)$ fundamental domain, $\CF=\BH/$SL$(2,\BZ)$. Taking $\CF$ as the familiar ``key
hole'' shaped region of the upper half plane extending along the imaginary axis, we take
$\BH/\Gamma^0(4)$ as the union of this domain and the images
under $\tau\to \tau+1$, $\tau+2$, $\tau+3$, $\tau+4$, $-1/\tau$ and
$2-1/\tau$. 
  
Integrals over $\CF$ of modular invariant
integrands, $d\tau \wedge d\bar \tau\, F$, 
are well-studied in the literature. They appear for
example as inner product on
the space of modular forms \cite{Petersson1950}, as one-loop amplitudes in string theory \cite{Dixon:1990pc, Harvey:1995fq, Lerche:1988np} or in the context of divisors on symmetric spaces
\cite{Borcherds:1996uda}. Depending on the integrand, different
techniques are available to evaluate the integral. A common approach to evaluate the integral is to ``unfold''
$\CF$ to the strip $\tau\in\mathbb{H}, \tau_1\in
[-\frac{1}{2},\frac{1}{2}]$, either using an Eisenstein or
Poincar\'e series in the integrand, or using the technique of lattice reduction
\cite{Dixon:1990pc,Harvey:1995fq,Borcherds:1996uda}. The latter was also
originally used for the $u$-plane integral \cite{Moore:1997pc}.

However, the integral may be evaluated more straightforwardly in special cases. Namely when the integrand $F$ can be expressed as
a total derivative with respect to $\bar \tau$, 
\be
\label{FdH}
F=\frac{\partial H}{\partial \bar \tau},
\ee
with $H=H(\tau,\bar \tau)$ a function which transforms as a modular form of weight
$2$.  As reviewed in
appendix \ref{int_fund_dom}, the integral receives in this case only a contribution from the cusp at
$ i\infty$, and the final result is
\be
\label{cuspinfty}
\int_{\CF} d\tau \wedge d\bar \tau\, F= \left[H\right]_{q^0}.
\ee

An integral $\Phi$ over the fundamental domain for $\Gamma^0(4)$,
$\BH/\Gamma^0(4)$, can similarly be expressed as a sum over its three
cusps at $i\infty$, $0$ and $2$:
\be
\label{Phiinf02}
\Phi=\Phi_\infty+\Phi_0+\Phi_2.
\ee 
The contribution of the cusp at infinity, $\Phi_\infty$, is
\be
\label{Phiinfty}
\Phi_\infty=4\,\left[H\right]_{q^0},
\ee
which differs by a factor $4$ from (\ref{cuspinfty}) since the arc for
large $\tau_2$ runs now from $\tau_1=3\frac{1}{2}$ to
$\tau_1=-\frac{1}{2}$. The contributions from the other two cusps, at $\tau=0$ and $2$, can
be mapped to $i\infty$ using the transformation $\tau\to -1/\tau$ and
$\tau\to 1/(2-\tau)$ respectively.

\subsection{The $u$-plane integrand as a total derivative}
 In order to evaluate the integral (\ref{Phi_final}), it is clear from
 the above  that we can readily evaluate the integral, if we may express the integrand as a total derivative. 
To this end, we need to find a non-holomorphic modular form $H$ of
weight two, which satisfies
\be
\label{dbartauH}
\partial_{\bar \tau}H=\tilde \nu\,\Psi_\bfmu^{J}.
\ee
In the remainder of this section, we determine such an $H$ making use
of indefinite theta series \cite{ZwegersThesis}. This allows us to
rederive the Donaldson invariants for Hirzebruch surfaces and the
projective plane, and also the wall-crossing formula for $\Phi_{\bfmu}^J$ for an arbitrary
simply connected four-manifold $M$ with $b_2^+=1$,
See Appendix \ref{Zwegers_theta} for a concise introduction to
indefinite theta functions. Here we recall the definition of $\widehat
\Theta^{JJ'}_{\bfmu}:\mathbb{H}\times \mathbb{C}^{b_2}\to
\mathbb{C}$. The function $\Theta^{JJ'}_{\bfmu}$ depends on two parameters $J$ and $J'$.  As the
notation suggests, we identify $J$ with the period point
of the metric of $M$ in $H^2(M)$. We choose furthermore $J'\in H^2(M,\BZ)$ such that
$(J')^2=0$ and $B(J,J')>0$. We deduce from the classification of
Lorentzian lattices, Eqs (\ref{Qodd}) and (\ref{Qeven}), that such a vector does indeed exist for any smooth,
closed, oriented four-manifold $M$ with $b_2^+=1$ and $b_2>1$. With this choice of
parameters, $\Theta^{JJ'}_{\bfmu}$ is defined as
\be 
\label{hatTheta_sec}
\begin{split}   
\widehat \Theta^{JJ'}_{\bfmu}\!(\tau,\bfz)=\sum_{\bfk\in \Lambda+\mu} &
\tfrac{1}{2}\left( E(\sqrt{2\tau_2}\,B(\bfk+\bfb, \underline J))-\sgn(\sqrt{2\tau_2}\,B(\bfk+\bfb,J'))\right)\\
& \times (-1)^{B(\bfk,K_M)} q^{-\bfk^2/2} e^{-2\pi i B(\bfz,\bfk)},
\end{split}  
\ee  
where $E(t):\mathbb{R}\to [-1,1]$ is a reparametrisation of the error function,
\begin{equation}
E(t) = 2\int_0^t e^{-\pi u^2}du = \text{Erf}(\sqrt{\pi}t),
\end{equation}
and $\underline J=J/\sqrt{Q(J)}$ is the normalisation of $J$ as before.

Appendix \ref{Zwegers_theta} discusses that $\widehat
 \Theta^{JJ'}_{\bfmu}$ transforms as a modular form of weight $b_2/2$
 under $\Gamma^0(4)$ (\ref{theta_comp_mod}), and that its derivative
 to $\bar \tau$ equals
\be
\label{shadow_sec}
\begin{split} 
\partial_{\bar \tau} \widehat
\Theta^{JJ'}_\bfmu(\tau,\bfz)=& \Psi^J_\bfmu(\tau,\bfz),
\end{split}
\ee
with $\Psi^J_\bfmu(\tau,\bfz)$ equal to the sum over U(1) fluxes
(\ref{Psi}). We see from (\ref{shadow_sec}) that if we set  
\be
\label{HTheta}
H=\tilde \nu(\tau)\,\widehat
\Theta^{JJ'}_{\bfmu}(\tau,\bfrho)\,e^{2pu+\bfx^2 G},
\ee
then it satisfies indeed (\ref{FdH}) with $F$ the integrand of
$u$-plane integral (\ref{Phi_final}).  Note that there is an ambiguity in (\ref{shadow_sec}) since
addition of a holomorphic modular form to $\widehat
\Theta^{JJ'}_\bfmu$ does not change the right hand side. This is
related to the ambiguity in the choice of $J'$. We assume that this
ambiguity can be fixed by basic arguments, for example the existence
of an ``empty'' chamber. We will see this later in this section for
the Hirzebruch surfaces and the projective plane $\mathbb{P}^2$. 

Recall from (\ref{Phiinf02})
that the contributions to the integral are coming from the three
cusps. The contribution from the cusp at infinity (\ref{Phiinfty}) is
given by
\be  
\label{cont_infty}
4 \left[   \tilde \nu(\tau)\,
  \Theta^{JJ'}_\bfmu(\tau,\bfrho)\, e^{2pu+\bfx^2 G}\right]_{q^0},
\ee 
where $\Theta^{JJ'}_\bfmu$ is the holomorphic indefinite theta
series (\ref{indeftheta}), obtained from (\ref{hatTheta_sec}) by replacing $E(x)$
by $\sgn(x)$. One first expands in (\ref{cont_infty}) in the fugacities $p$ and $\bfx$, and then in $q$.
The contributions to the integral from the other cusps follows
similarly after transforming $\tau \to -1/\tau$ and $1/(2-\tau)$ in
the integrand. If the metric of $M$ has positive scalar curvature, the $u-$plane integral completely determines the Donaldson invariants. 

We briefly mention the wall-crossing formula which was earlier derived from the $u$-plane
integral in \cite{Moore:1997pc}. This formula gives the discontinuous change of
$\Phi_{\bfmu}^J$ under the variation of a metric with period point $J_0$ to one with period point $J_1 \in H^2(M)$. It is clear from the above that the difference $\Delta \Phi^{J_1J_0}_{\bfmu}=\Phi_{\bfmu}^{J_{1}}-\Phi_{\bfmu}^{J_{0}}$ is given by
$$
\Delta\Phi_{\bfmu}^{J_1J_0}(p,\bfx)= \int_{\mathbb{H}/\Gamma^0(4)}
d\tau\wedge d\bar \tau\,\tilde \nu
\left(\Psi_\bfmu^{J_1}-\Psi_\bfmu^{J_0}\right) e^{2pu+\bfx^2 G}.
$$
The contribution from the cusp at $i\infty$ gives then
\be
\Delta \Phi_{\bfmu}^{J_1J_0}(p,\bfx)=4\,\left[ \tilde \nu(\tau)\,\Theta^{J_1J_0}_\bfmu(\tau,\bfrho) \,e^{2pu+\bfx^2G}\right]_{q^0},
\ee 
while the contributions of other cusps are cancelled the wall-crossing
of Seiberg-Witten invariants \cite{Moore:1997pc}. This reproduces G\"ottsche's wall-crossing formula \cite[Theorem
3.3]{Gottsche:1996} and the expression of G\"ottsche-Zagier in terms
of an indefinite theta series \cite[Corollary 4.3]{Gottsche:1996aoa}.

\subsection{Application to the Hirzebruch surfaces $\BF_\ell$}
In this subsection, we specialize the four-manifold $M$ to one of the Hirzebruch surfaces
$\BF_{\ell}$. A Hirzebruch surface is a fibration $\pi:\BF_{\ell}\to \bfC$ with fibre $\bff\cong \BP^1$ over a base
$\bfC\cong \BP^1$. The base and the fibre form a basis for
$H^2(\BF_\ell,\BZ)$, in terms of which the canonical class $K_{\ell}$ is
expressed as $K_\ell=-2\bfC-(2+\ell) \bff$. The intersection matrix for $(\bfC,\bff)$ is 
\be
\label{intHirz} 
Q_{\BF_\ell}=\left( \begin{array}{cc} -\ell & 1 \\ 1 & 0 \end{array}\right).   
\ee
Note in particular that $\bff$ is an element of $H^2(\BF_\ell)$ with vanishing norm, $\bff^2=0$.
Two Hirzebruch surfaces $\BF_{\ell_1}$ and $\BF_{\ell_2}$ are (real)
diffeomorphic if $\ell_1=\ell_2 \mod 2$, while they are complex diffeomorphic
only for $\ell_1=\ell_2$. For more details on Hirzebruch surfaces see
for example \cite{Barth}.

To evaluate $\Phi_\bfmu^J$ for $\BF_{\ell}$ using (\ref{HTheta}),
consider the indefinite theta function (\ref{hatTheta_sec}) with the
quadratic form (\ref{intHirz}) above. We set $J'=\bff$, which is fixed
by the fact that no stable bundles exist for metrics with this period
point. Indeed for $J=\bff$, $\Theta_{\bfmu}^{J\bff}$ vanishes. One may
show that only the cusp at $\infty$ contributes to the integral for
$\BF_\ell$, and we arrive thus for $\Phi_{\bfmu}^J$ at the following
expression 
\be
\Phi_{\bfmu}^J(p,\bfx)=32i\left[ (u^2-1)\,\frac{da}{du}\,
  \Theta^{J\bff}_\bfmu(\tau,\bfrho)\, e^{2pu+\bfx^2 G}\right]_{q^0}.
\ee

We can simplify the expression for $\Theta^{J\bff}_{\bfmu}$ and
express it as a (generalized) Appell sum. To
this end, write $\bfk$ as $\bfk=\bfm + n\bff$, with $\bfm$
such that 
$$B(\bfm+\bfb,J)/B(\bff,J)\in [0,1).$$ 
Then $\Theta_{\bfmu}^{J\bff}$ takes the form
\be
\begin{split}
\Theta_{\bfmu}^{J\bff}(\tau,\bfrho)=&\sum_{\bfm
    \in \Lambda+\bfmu\atop B(\bfm+\bfb,J)/B(\bff,J)\in
    [0,1)}\sum_{n\in\BZ} \,\,\, (-1)^{B(\bfm, K_\ell)} q^{-\bfm^2/2} e^{-2\pi i B(\bfrho,\bfm) }\\
& \times  \tfrac{1}{2}\left(\sgn(B(\bfm+\bfb,J)+n B(\bff,J)) - \sgn(B(\bfm+\bfb, \bff))  \right)  \\
&\,\,\,\,\,\,\, \times (-1)^{nB(\bff, K_\ell)} q^{-nB(\bff,\bfm)} e^{-2\pi i n
  B(\bfrho,\bff) }.
\end{split}
\ee
Carrying out the sum over $n$ as a geometric series, we find
\be
\begin{split}
\Theta_{\bfmu}^{J\bff}(\tau,\bfrho)=&\sum_{\bfm
    \in \Lambda+\bfmu\atop B(\bfm+\bfb,J)/B(\bff,J)\in
    [0,1)} \frac{(-1)^{B(\bfm, K_\ell)} q^{-\bfm^2/2} e^{-2\pi i B(\bfrho,\bfm) }}{1-q^{-B(\bff,\bfm)}e^{-2\pi i B(\bfrho,\bff)}},
\end{split}
\ee 
where we used that $B(\bff, K_\ell)=-2$. 

Of particular interest in the literature is the suitable
polarization $J_{\eps}=(\eps (\bfC+\ell \bff)+\bff)/\sqrt{\ell
  \eps^2+\eps}$, with $\eps$ sufficiently small such that no walls are
crossed between $\bff$ and $J_\eps$ for the rank two vector bundles. If
$B(\bfmu,\bff)\in \mathbb{Z}+\frac{1}{2}$ the condition on
$\bfm$ has no solutions in agreement with the fact there are no stable
bundles for such metrics. If  $B(\bfmu,\bff)\in \mathbb{Z}$ we
have the solutions $\bfm=0$ and $\bfm=\frac{1}{2}f$, due to strictly
semi-stable bundles. One finds therefore
\be
\begin{split}
&\Theta_{0}^{J_\eps \bff}(\tau,\bfrho)=\frac{1}{1-e^{-2\pi i B(\bfrho,\bff)}},\\
&\Theta_{\bff}^{J_\eps J'}(\tau,\bfrho)=\frac{- e^{-\pi i B(\bfrho,\bff)
  }}{1-e^{-2\pi i B(\bfrho,\bff)}}=\frac{i}{2 \sin(\pi B(\bfrho,\bff))}.
\end{split}
\ee
Using (\ref{nutau0}) and letting $\bfx=x_C\bfC+x_f \bff\in H^2(\BF_\ell,\mathbb{R})$, we arrive
at the following non-vanishing generating function for Donaldson
invariants for the suitable polarization $J_{\eps}$,
\be 
\begin{split}
&\Phi^{J_\eps}_{0}(p,\bfx)=-16\left[ (u^2-1) \frac{da}{du} e^{2p u + \bfx^2 G(u)}   \cot(\tfrac{1}{2} x_C\, du/da )   \right]_{q^0},\\
&\Phi^{J_\eps}_{\frac{1}{2}\bff}(p,\bfx)=16\left[ (u^2-1) \frac{da}{du} e^{2p u + \bfx^2 G(u)}  \frac{1}{\sin(\tfrac{1}{2} x_C\, du/da )}   \right]_{q^0},
\end{split}
\ee
where we expressed $\Phi^{J_\eps}_{0}$ in terms of $\cot(x)$ using the
fact that only odd powers of $x_f$ contribute to the expansion of the right hand side. This is in agreement with \cite[Theorem 5.3]{Gottsche:1996aoa} and \cite[Section 8.2]{Moore:1997pc}.

\subsection{Application to the projective plane $\BP^2$}
We consider  the complex projective plane $\BP^2$ as another
application of indefinite theta functions to the $u$-plane
integral. Since $b_2(\BP^2)=1$ in this case, the period point of the
metric is proportional to the hyperplane class $H$. Since there is
thus no chamber dependence, we omit it from the notation. The sum over
U(1) fluxes $\Psi_\mu$ is given by\footnote{We omit the boldface font
  here for $k$, $b$ and $\rho$, since they are elements of one-dimensional spaces.
}
\be
\begin{split}
\Psi_\mu(\tau,\rho)=&\exp\!\left( -2\pi\tau_2 b^2\right) \sum_{k\in \BZ+\mu} \partial_{\bar \tau}\left(\sqrt{2\tau_2}(k+ b)\right)\,\\
&\times (-1)^{3k}\,\bar q^{k^2/2} e^{-2\pi i \bar \rho k},
\end{split}
\ee
where we have used that the canonical class $K_{\BP^2}$ equals $3H$.

Since the lattice $H^2(\BP^2,\BZ)$ is one-dimensional, we can not directly apply the indefinite theta function to integrate over the Coulomb branch. 
However, we can extend the one-dimensional lattice to a two-dimensional
lattice by dividing and multiplying by the Jacobi theta function
$\theta_4$ defined in (\ref{Jacobitheta}): $\Psi_\mu=\frac{\theta_4(\tau)}{\theta_4(\tau) }\Psi_\mu$. Geometrically one may interpret these
manipulations in terms of the blow-up $\widehat \BP^2$ of $\BP^2$;
note that the measure (\ref{tildenu}) differs by a factor
$\theta_4^{-1}$ for $\BP^2$ and $\widehat \BP^2$. Including the
summation over $\BZ$ in $\theta_4$ in the lattice sum, $\Psi_\mu(\tau,\rho)$ reads
\begin{equation}
\label{ththPsi}
\begin{split}
\Psi_\mu(\tau,\rho)=&\frac{\exp\!\left( -2\pi \tau_2b^2 \right)}{\theta_4(\tau)}\\
&\times \sum_{(k_1,k_2)\in \mathbb{Z}^2+(\mu,0)}
\partial_{\bar \tau}\! \left(\sqrt{2\tau_2}(k_1+b)\right)\,(-1)^{3k_1+k_2}\,\bar
q^{k_1^2/2} q^{k_2^2/2} e^{-2\pi i \bar \rho k_1}.
\end{split}
\end{equation}

Our earlier discussion shows that $\Psi_{\mu}(\tau,\rho)$ can be
expressed as an anti-holomorphic derivative,
\be
\frac{1}{\theta_4(\tau)}\partial_{\bar \tau} \widehat 
\Theta^{JJ'}_\mu(\tau,\rho),
\ee
where $\widehat \Theta^{JJ'}_\bfmu$ is the completion of the
indefinite theta function $\Theta^{JJ'}_\bfmu$ whose associated
lattice $\Lambda$ is the two-dimensional lattice with diagonal
quadratic form $\textrm{diag}(1,-1)$. The two-dimensional parameters
$\bfmu$ and $\bfrho$ are given by $(\mu,0)$ and $(\rho,0)$
respectively, whereas the two parameters $J,J'\in \Lambda\otimes \BR$
are given by $J=(1,0)$ and $J'=(1,1)$ respectively. 

The lattice sum in $\Theta^{JJ'}_\bfmu$ can be partially carried out
using a geometric series, leading to the expression 
\be
\label{appell}
\Theta^{JJ'}_\mu(\tau,\rho)=w^\mu(-1)^{2\mu}\sum_{\ell\in \BZ+\mu}
\frac{(-1)^\ell q^{\frac{1}{2}\ell^2+\mu\ell}}{1-wq^\ell},
\ee
with $w=e^{2\pi i \rho}$. This is, up to a prefactor, a specialization of the
Appell sum \cite{ZwegersThesis}
\be
\label{Appell}
A(u,v,\tau)=e^{\pi i u}\sum_{n\in
  \BZ}\frac{(-1)^nq^{n(n+1)/2}e^{2\pi i nv}}{1-e^{2\pi i u}q^n}.\ee
 
Treating first the case $\mu=\frac{1}{2}$, we arrive at the following expression for the generating function $\Phi_{\frac{1}{2}}(p,x)$ 
\be
\Phi_{\frac{1}{2}}(p,x)=-32i\left[ 
(u^2-1)\,\frac{da}{du}\,e^{2pu+x^2 G(u)}\,
 \Theta_{(\frac{1}{2},0)}^{JJ'}\!\left(\tau,\rho\right)\right]_{q^0},
\ee
which gives for the first few terms
 \be
\Phi_{\frac{1}{2}}(0,x)=1+\frac{3}{16}\frac{x^4}{4!}+\frac{29}{32}\frac{x^8}{8!}+\frac{69525}{4096}\frac{x^{12}}{12!}+O(x^{16}).
\ee
These terms are in agreement with \cite[Theorem
4.4]{ellingsrud1995wall}, while the full
series matches the expression of G\"ottsche \cite[Theorem 3.5]{Gottsche:1996}.

Next we consider $\mu=0$. The series $\Phi_0(p,x)$ can be determined similarly using multiplication and
division by $\theta_4$. However, we notice from (\ref{appell}) that
$\widehat \Theta_0(\tau,\rho)$ is then divergent for small $\rho$, which is
at odds with the Donaldson invariants being polynomials in $\bfx$. The
resolution is that the holomorphic integration constant mentioned
below (\ref{FG}) is non-vanishing in this case. Using the blow-up formula, one finds that the constant equals
$$ 
C(\tau,\rho)=\frac{\theta_4(\tau,0)}{\theta_4(\tau,\rho)}\,\partial_\rho
\ln\!\left(\frac{\theta_1(\tau,\rho)}{\theta_4(\tau,\rho)} \right), 
$$
leading to the following expression for $\Phi_0(p,x)$ 
\be
\label{Phi1}
\Phi_0(p,x)=-32i\left[ 
(u^2-1)\,\frac{da}{du}\,e^{2pu+x^2 G(u)}\,\left(C(\tau,\rho)+ 
 e^{-\pi i \rho}\,A(\rho,-\tfrac{1}{2}\tau,\tau)\right)\right]_{q^0}. 
\ee 

To relate this to the expression of \cite[Theorem
3.5]{Gottsche:1996}. We recall the periodicity property of the Appell
function (\ref{Appell}) from \cite[Chapter 1]{ZwegersThesis},
\be
\frac{A(u,v,\tau)}{\theta_1(v,\tau)}-\frac{A(u+z,v+z,\tau)}{\theta_1(v+z,\tau)}=\frac{\eta^3\,\theta_1(u+v+z,\tau)\,\theta_1(z,\tau)}{\theta_1(u,\tau)\,\theta_1(v,\tau)\,\theta_1(u+z,\tau)\,\theta_1(v+z,\tau)}.
\ee
Letting $v=-\frac{1}{2}\tau$ and taking the limit $z\to
\frac{1}{2}\tau$, one finds
\be
\label{C+A}
\begin{split} 
C(\tau,\rho)+ e^{-\pi i \rho}\,A(\rho,-\tfrac{1}{2}\tau,\tau)= &\frac{\theta_4(\tau)}{\eta(\tau)^3}\sum_{k_1\in \BZ\atop k_2\in \BZ+\frac{1}{2}} \left(\sgn(k_1+a)-\sgn(k_1+k_2+a) \right)\\
&\times k_2\, (-1)^{k_1+k_2} e^{2\pi i \rho k_1}q^{-k_1^2/2+k_2^2/2}.
\end{split}
\ee 
Substitution of this expression in (\ref{Phi1}) reproduces the
expression in \cite[Theorem 3.5]{Gottsche:1996}. For completeness, we
list the first few terms in the expansion 
\be
\Phi_0(p,x)=-\frac{3}{2}x+\frac{x^5}{5!}+3\frac{x^9}{9!}+54\frac{x^{13}}{13!}+O(x^{17}),
\ee
in agreement with \cite[Theorem
4.2]{ellingsrud1995wall} and  \cite{Moore:1997pc}. One can
arrive at the rhs of (\ref{C+A}) alternatively by multiplying and dividing in (\ref{ththPsi}) by $\theta_1(z,\tau)$
rather than $\theta_4(\tau)$, and taking the limit $z\to 0$ in
$\Theta^{JJ'}_0(\tau,\rho,z)/\theta_1(z,\tau)$. A similar procedure was used
in the context of D3-instanton corrections \cite[Section 4]{Alexandrov:2017qhn}.

      
\section{On the extension to gauge groups with rank $>1$} \label{higherrank}
Donaldson-Witten theory can be generalized to theories with a gauge
group $G$ with rank $r>1$ \cite{LoNeSha, Marino:1998bm} using the corresponding Seiberg-Witten geometries \cite{Klemm:1994qs, Argyres:1994xh,Douglas:1995nw, Tachikawa:2011yr}. Mochiziku \cite{Mochiziku:2009} developed an algebraic-geometric framework to discuss higher rank Donaldson invariants. This section generalizes the $\mathcal{\bar  Q}$-exact surface operator (\ref{I+0}) to theories with  gauge group $G$, and discusses the sum over U$(1)^r$ fluxes of the Coulomb branch integrand for a four-manifold $M$ with $b_1=0$ and $b_2^+=1$. We keep this section relatively short and refer the reader for the details to \cite{Marino:1998bm}. 
 
Let us consider the Coulomb branch of a $\mathcal{N}=2$ supersymmetric Yang-Mills theory whose gauge group $G$ has rank $r$. We denote the Cartan elements of the Lie algebra by $H_K$, $K=1, \ldots, r$. Then, the vacuum expectation value of the scalar component of the $\CN=1$ chiral superfield can classically be brought to the form 
\be
\phi = \sum_{K=1}^r a^K H_K.
\ee  
The $a^K$ provide local special coordinates on the Coulomb branch
moduli space. Alternatively, one can consider the $r$ Weyl invariant
Casimirs $u_K$, $K=1,\dots, r$, as coordinates on the Coulomb
branch. At a generic point on the Coulomb branch, the field content
consists of $r$ copies of the effective U(1) theory described in
Section \ref{SectionSW}, which are distinguished by a superscript:
$A^K$, $\psi^K$, $\dots$, for $K=1,\dots, r$. The effective coupling
$\tau_{KL}=\tau_{KL}(a^M)$ is now an $r\times r$ matrix. The effective
Coulomb branch theory breaks down at the locus where gauge bosons
become massless, or more geometrically, the corresponding
Seiberg-Witten curve becomes singular. 

Most aspects of the rank one Donaldson-Witten theory generalize to rank $r$ without much effort. For example after topological twisting, the action of the $\mathcal{\bar Q}$ operator on the fields is given by
\be 
\label{barQcomm2}
\begin{split}
&[\mathcal{\bar Q},A^K]=\psi^K, \hspace{55pt} [\mathcal{\bar Q},a^K]=0, \hspace{50pt} [\mathcal{\bar Q},\bar a^K]=\sqrt{2}i \eta^K,  \\
&   [\mathcal{\bar Q},D^K]=(d_A\psi^K)_+, \qquad  \{\mathcal{\bar Q},\psi^K \}=4\sqrt{2}\,da^K, \\
& \{\mathcal{\bar Q},\eta^K \}=0, \hspace{58pt}    \{\mathcal{\bar Q},\chi^K\}=i(F_+-D)^K.
\end{split}
\ee 
The effective Lagrangian on the Coulomb branch is similarly a
straightforward generalization of the rank 1 case
\cite{Marino:1998bm}.

There is a larger freedom for the construction of surface operators in the higher rank theories. Starting from any invariant function $\CU=\CU(a^K)$ of the coordinates $a^K$, one may construct a suitable surface operator. The operator $\tilde I_-$ (\ref{I-}) generalized to general $r$ takes the form \cite{Marino:1998bm}
\[
\tilde{I}_{-}(\bfx) = \int_{\bfx} \frac{i}{32\sqrt{2}\pi} \CU_{KL} \psi^K \wedge \psi^L - \frac{i}{4\pi} \CU_K (F_{-}+ D)^K,
\]
where the subscripts indicate differentiation to $a^K$:
\[
{\CU}_K = \frac{d\,\mathcal{U}}{ d {a}^K},\qquad {\CU}_{KL} = \frac{d^2\,\mathcal{U}}{ d {a}^K d {a}^L}.
\]
The generalization of the $\bar \CQ$-exact surface operator $\tilde{I}_+$ (\ref{I+}) is similarly given by
\be
\tilde{I}_{+}(\bfx) =  -\frac{1}{4\pi} \int_{\bfx} \left\{ \bar{\CQ},  \bar{\CU}_K\, \chi^K \right\},
\ee
which using the algebra (\ref{barQcomm2}) becomes
\be
\tilde{I}_{+}(\bfx) = -\frac{i}{2\sqrt{2}\pi} \int_{\bfx}  \bar{\CU}_{KL} \eta^K \chi^L + \frac{1}{\sqrt{2}}\, \bar \CU_K (F_{+} - D)^K.
\ee

Our next aim is derive the sum over the $U(1)^r$ fluxes $\bfk^K$,
$\Psi^J_{r,\mu}$, when both $\tilde I_+$ and $\tilde I_-$ are inserted
in the path integral. After integrating out the auxiliary fields $D^K$, we find that $\Psi^J_{r,\mu}$ is given by
\be
\label{Psir}
\begin{split}
&\Psi^J_{r,\mu}(\tau_{KL},\bfrho_K)=\frac{1}{\sqrt{\text{det}\, v}}\, e^{ -2\pi v_{KL} \bfb_+^K\bfb_+^L} \sum_{\bfk\in \Gamma}(-1)^{B(\bfk^K W_K , K_{M})} \,\CK(\bfk, \bfrho, \omega) \\
  &\times \exp\!\Big(\!-i\pi \bar{\tau}_{KL}B(\bfk_{+}^K , \bfk_{+}^L) - i\pi \tau_{KL}B(\bfk_{-}^K, \bfk_{-}^L) - 2\pi i B(\bfk_{+}^K , \bar{\bfrho}_K) - 2\pi iB( \bfk_{-}^K , \bfrho_K) \Big) 
\end{split}
\ee   
where $v_{KL}=\mathrm{Im}(\tau_{KL})$, $W_K$ are the components of the Weyl vector of $G$, and we introduced 
\[
\bfrho_K \equiv \frac{\bfx}{2\pi}\, \CU_K \in H^2(M,\BC),\qquad \bfb^K = v^{KL} \text{Im}(\bfrho_L)\in H^2(M,\BR),
\] 
in analogy to the rank one case. The kernel $\CK$ in (\ref{Psir}) is given by the integral over the fermion zero modes
\be
\begin{split}
&\CK(\bfk,\bfrho,\omega)= \int \left[ \prod_{K,L=1}^r  d\eta_0^K d\chi_0^L  \right] \exp\! \Big( -\frac{\sqrt{2}i}{4}\int_M  \bar{\CF}_{KLM}  \eta_0^K \chi_0^L \wedge (\bfk_+-\bfb_+)^M  \\
& - \frac{i}{\sqrt{2}}  \bar{\bfrho}_{KL} \eta_0^K \chi_0^L    +  \frac{1}{64\pi} \,v^{KP}   \bar{\CF}_{KLM}   \bar{\CF}_{PQR}  \eta_0^L \chi_0^M\wedge \eta_0^Q \chi_0^R \Big),
\end{split}
\ee
where $\CF_{KLM}=d\tau_{KL}/da^M$. Carrying out this integral for $G=$SU$(3)$ ($r=2$), we arrive at
\be
\non
\begin{split}
\CK(\bfk,\bfrho,\omega)&= \tfrac{1}{8}\left( \bar{\CF}_{11K}
  B(\bfk^K-\bfb^K,J)  + 2B(\bar \bfrho_{11}, J) \right) \left(\bar{\CF}_{22L}   B(\bfk^L-\bfb^L,J)  + 2B(\bar \bfrho_{22},J)  \right) \\
&- \tfrac{1}{8}\left( \bar{\CF}_{12K}   B(\bfk^K-\bfb^K,J)  +2 B(\bar \bfrho_{12},J) \right) \left( \bar{\CF}_{12L}  B(\bfk^L-\bfb^L,J)+  2 B(\bar \bfrho_{12},J) \right) \\
& + \frac{1}{32\pi}(\bar{\CF}_{11K}\bar{\CF}_{22L}- \bar{\CF}_{12K}\bar{\CF}_{12L})v^{KL}.
\end{split}
\ee
We leave it for future work to express $\Psi^J_{r,\mu}$ as a total derivative of $\bar a^K$.


\section{Conclusion and discussion}  
We have discussed partition functions of Donaldson-Witten theory for a four-manifold $M$ with $b_2^+=1$,
and in particular the contribution of the integral over the Coulomb
branch. We have shown that for gauge group SU$(2)$ and SO(3) the integrand may be expressed as $\bar 
\tau$-derivative of an indefinite theta function, after insertion of a
$\mathcal{\bar Q}$-exact surface operator $\tilde I_+$ which couples to the
self-dual part of the field strength $F$. This allows to
readily evaluate the integral, and to express it as a sum over the cusps of the
integration domain. In this way, we reproduce the result of G\"ottsche, who expressed generating series of
Donaldson invariants in terms of a residue of an indefinite theta
function. 

There are various directions to which our results may be applied and
generalized, in particular the evaluation of partition
functions of other four-dimensional theories, such as those including matter and higher rank gauge groups. Besides the fundamental interest
in path integrals of Yang-Mills theories, this may also prove useful for establishing new
four-manifold invariants \cite{Marino:1998tb}. \\

\noindent {\bf Acknowledgements}\\
We would like to thank Gregory Moore and Samson Shatashvili for useful
discussions and correspondence. JM thanks IHES for hospitality during the completion of this work.


\appendix
\section{Modular forms and theta functions}
\label{app_mod_forms}
We collect in this appendix a few essential aspects of the theory of
modular forms. For more comprehensive treatments we refer the reader
to the available literature. See for example \cite{Serre,Zagier92,Bruinier08}. 

\subsection*{Modular groups}
The modular group $\operatorname{SL}(2,\mathbb{Z})$ is the group of integer matrices
with unit determinant
\be
\operatorname{SL}(2,\mathbb{Z})=\left\{ \left. \left( \begin{array}{ccc}
a & b   \\
c & d  \end{array} \right) \right| a,b,c,d\in \BZ; \, ad-bc=1\right\}.
\ee
We introduce moreover the congruence subgroup $\Gamma^0(n)$ 
\begin{equation}
\label{Gamma04} 
\Gamma^0(n) = \left\{  \left.\left( \begin{array}{ccc}
a & b   \\
c & d  \end{array} \right) \in \text{SL}(2,\BZ) \right| b = 0 \text{ mod } n \right\}.
\end{equation}

\subsection*{Eisenstein series}
We let $\tau\in \mathbb{H}$ and define $q=e^{2\pi i \tau}$. Then the Eisenstein series $E_k:\mathbb{H}\to \mathbb{C}$ for even $k\geq 2$ are defined as the $q$-series 
\be
\label{Ek}
E_{k}(\tau)=1-\frac{2k}{B_k}\sum_{n=1}^\infty \sigma_{k-1}(n)\,q^n,
\ee
with $\sigma_k(n)=\sum_{d|n} d^k$ the divisor sum. For $k\geq 4$, $E_{k}$ is a modular form of
$\operatorname{SL}(2,\mathbb{Z})$ of weight $k$. In other words, it transforms under $\operatorname{SL}(2,\mathbb{Z})$ as
\be
E_k\!\left( \frac{a\tau+b}{c\tau+d}\right)=(c\tau+d)^kE_k(\tau).
\ee
On the other hand $E_2$ is a quasi-modular form, which means that the $\operatorname{SL}(2,\mathbb{Z})$ transformation of $E_2$ includes a shift in addition to the weight,
\be 
\label{E2trafo}
E_2\!\left(\frac{a\tau+b}{c\tau+d}\right) =(c\tau+d)^2E_2(\tau)-\frac{6i}{\pi}c(c\tau+d).
\ee

\subsection*{Dedekind eta function}
The Dedekind eta function $\eta:\mathbb{H}\to\mathbb{C}$ is defined as
\be
\eta(\tau)=q^{\frac{1}{24}}\prod_{n=1}^\infty (1-q^n).
\ee
It is a modular form of weight $\frac{1}{2}$ under SL$(2,\BZ)$ with a
non-trivial multiplier system. It transforms under the generators of
SL$(2,\BZ)$ as
\be
\begin{split}
&\eta(-1/\tau)=-i\sqrt{-i\tau}\,\eta(\tau),\\
&\eta(\tau+1)=e^{\frac{\pi i}{12}}\, \eta(\tau). 
\end{split}
\ee

\subsection*{Jacobi theta functions}
The four Jacobi theta functions $\vartheta_j:\mathbb{H}\times
\mathbb{C}\to \mathbb{C}$, $j=1,\dots,4$, are defined as
\be
\label{Jacobitheta}
\begin{split}
&\vartheta_1(\tau,v)=i \sum_{r\in
  \mathbb{Z}+\frac12}(-1)^{r-\frac12}q^{r^2/2}e^{2\pi i
  rv}, \\
&\vartheta_2(\tau,v)= \sum_{r\in
  \mathbb{Z}+\frac12}q^{r^2/2}e^{2\pi i
  rv},\\
&\vartheta_3(\tau,v)= \sum_{n\in
  \mathbb{Z}}q^{n^2/2}e^{2\pi i
  n v},\\
&\vartheta_4(\tau,v)= \sum_{n\in 
  \mathbb{Z}} (-1)^nq^{n^2/2}e^{2\pi i
  n v}. 
\end{split}
\ee

We let $\vartheta_j(\tau,0)=\vartheta_j(\tau)$ for $j=2,3,4$.
Their transformations under the generators of $\Gamma^0(4)$ are 
\be
\label{Jacobitheta_trafos}
\begin{split}
&\vartheta_2(\tau+4)=-\vartheta_2(\tau),\qquad
\vartheta_2\!\left(\frac{\tau}{\tau+1}\right)=\sqrt{\tau+1}\,\vartheta_3(\tau),  \\
&\vartheta_3(\tau+4)=\vartheta_3(\tau),\qquad
\vartheta_3\!\left(\frac{\tau}{\tau+1}\right)=\sqrt{\tau+1}\,\vartheta_2(\tau),  \\
&\vartheta_4(\tau+4)=\vartheta_4(\tau),\qquad
\vartheta_4\!\left(\frac{\tau}{\tau+1}\right)=e^{-\frac{\pi
    i}{4}}\sqrt{\tau+1}\,\vartheta_4(\tau).  \\
\end{split}
\ee

\subsection*{Siegel-Narain theta function}
Siegel-Narain theta functions form a large class of theta functions of
which the Jacobi theta functions are a special case. We restrict here
to a specific Siegel-Narain theta function for which the associated lattice $\Lambda$
is a uni-modular lattice of signature
$(1,n-1)$ (or a Lorentzian lattice). We denote the bilinear form by
$B(\bfx,\bfy)$ and the quadratic form $B(\bfx,\bfx)\equiv Q(\bfx)\equiv \bfx^2 $.
Let $K$ be a characteristic vector of $\Lambda$, such that
$ Q(\bfk) + B(\bfk, K) \in 2\mathbb{Z}$ for each $\bfk\in \Lambda$.   

Given an element $J\in \Lambda\otimes \mathbb{R}$ with $Q(J)>0$, we
may decompose the space $\Lambda\otimes \mathbb{R}$ in a positive
definite subspace $\Lambda_+$ spanned by $J$, and a negative definite
subspace $\Lambda_-$, orthogonal to $\Lambda_+$. Let  $\underline J=J/\sqrt{Q(J)}$
be the normalization of $J$. The projections of a
vector $\bfk\in \Lambda$ to $\Lambda_+$ and $\Lambda_-$ are then given by
\be
\label{k+k-}
\bfk_+=B(\bfk,\underline J)\, \underline J, \qquad \qquad \bfk_{-} = \bfk-\bfk_+.
\ee

Given this notation, we can introduce the Siegel-Narain theta function
of our interest $\Psi^J_\bfmu:\mathbb{H}\times \mathbb{C}\to \BC$. Let $J$ be
as discussed above (\ref{k+k-}) and $\bfmu\in \Lambda\otimes
\mathbb{R}$. Then $\Psi^J_\bfmu$ is defined by\footnote{For brevity we
list in $\Psi^J_\bfmu$ only the holomorphic arguments $\tau$ and $z$,
even though the function does also depend on $\bar \tau$ and $\bar z$.}
\be 
\label{PsiJ} 
\begin{split} 
\Psi^J_\bfmu(\tau,\bfz)=&e^{-2\pi\tau_2 \bfb_+^2} \sum_{\bfk\in
  \Lambda + \bfmu} \partial_{\bar \tau} (\sqrt{2\tau_2}B(\bfk+\bfb,
\underline J))\,(-1)^{B(\bfk, K)} q^{-\bfk_-^2/2} \bar q^{\bfk_+^2/2} \\
&\times e^{-2\pi i B(\bfz, \bfk_-)-2\pi i B(\bar \bfz,\bfk_+)},
\end{split}
\ee
where $\bfb=\mathrm{Im}(\bfz)/\tau_2\in \Lambda \otimes \mathbb{R}$. The parameter $\bfb$ is typically taken independent of $\bar \tau$ in
the literature, and in that case (\ref{PsiJ}) simplifies. In the
application in the main text, $\bfz$ is actually a modular form of
weight $-1$, such that $\bfb$ is not independent of
$\bar \tau$. The derivative $\partial_{\bar
  \tau}\bfb$ transforms then as a modular form of mixed weight $(1,2)$. 

To determine the modular properties of $\Psi^J_\bfmu$, one may use the
standard technique of Poisson resummation, as for example in \cite{Borcherds:1996uda}. To this end, it is most convenient  to shift $\bfmu$ by $K/2$. 
One finds for the modular transformations of $\Psi^J_\bfmu$ under the generators of
SL$(2,\mathbb{Z})$ the following identities
\be
\label{Psi_trafos}
\begin{split}
\Psi^J_{\bfmu+K/2}(\tau+1,\bfz)=&e^{\pi i (\bfmu^2-K^2/4)}\,\Psi^J_{\bfmu+K/2}(\tau,\bfz+\bfmu), \\
\Psi^J_{\bfmu+K/2}(-1/\tau,\bfz/\tau)=& 
-i(-i\tau)^{\frac{n}{2}}(i\bar \tau)^{2} \exp\!\left(-\pi
  i \bfz^2/\tau+\pi i K^2/2\right) (-1)^{B(\bfmu,K)}\\
&\times  \Psi^J_{K/2}(\tau,\bfz-\bfmu).
\end{split}
\ee
We note that a shift in $\bfz$ by $\bfnu\in \Lambda \otimes
\mathbb{R}$ times $\tau$, can be related to a shift
in $\bfmu$ by the following identity
\be
\label{shiftz}
\Psi^J_\bfmu(\tau,\bfz+\bfnu \tau)=e^{2\pi i B(\bfz,\bfnu)}q^{\bfnu^2/2}
(-1)^{-B(\bfnu, K)} \Psi^J_{\bfmu+\bfnu}(\tau,\bfz),
\ee
while shifting $\bfz$ by $\bfnu\in \Lambda \otimes
\mathbb{R}$ gives
\be
\Psi^J_\bfmu(\tau,\bfz+\bfnu)=e^{-2\pi i B(\bfnu,\bfmu)} \Psi^J_{\bfmu}(\tau,\bfz).
\ee
 
Due to the relation (\ref{shiftz}), the parameters $\bfmu$ and $\bfz$ are somewhat
redundant as arguments of $\Psi^J_{\bfmu}$. They play however different roles in the
main part of this article. There $\bfmu$ is one half the second
Stiefel-Whitney class of a line bundle and therefore restricted to
$\Lambda/2$, while $\bfz$ is a fugacity valued in $\Lambda\otimes
\BC$. For such $\bfmu\in \Lambda/2$ one can show that $\Psi^J_\bfmu$ is a
modular form of the congruence subgroup $\Gamma^0(4)$. The  action of
the generators of $\Gamma^0(4)$ on $\Psi^J_\bfmu$ with $\bfmu\in \Lambda/2$ is given by
\begin{eqnarray} 
&&\label{Psi-1} \Psi^J_\bfmu(\tau,-\bfz)=-e^{2\pi i B(\bfmu,K)}\,\Psi^J_\bfmu(\tau,\bfz), \\
&&\label{PsiS}\Psi^J_\bfmu\!\left( \frac{\tau}{\tau+1}, \frac{\bfz}{\tau+1} \right)=(\tau+1)^{\frac{n}{2}}(\bar \tau+1)^2\exp\!\left(-\frac{\pi i\bfz^2}{\tau+1} +\frac{\pi i}{4}K^2\right) \Psi^J_{\bfmu}(\tau,\bfz),\\
&&\label{PsiT4}\Psi^J_\bfmu(\tau+4,\bfz)=e^{2\pi i B(\bfmu,K)}\,\Psi_\bfmu(\tau,\bfz).
\end{eqnarray}

\section{Indefinite theta functions for uni-modular lattices of
  signature $(1,n-1)$}
\label{Zwegers_theta} 
We discuss in this appendix various aspects of indefinite theta
functions and their modular completion. We assume that the associated
lattice $\Lambda$ is unimodular and of signature
$(1,n-1)$ and use the notation discussed in Appendix \ref{app_mod_forms}. 
 
To define the indefinite theta function, we choose two positive
definite vectors $J$ and $J'\in \Lambda\otimes \mathbb{R}$ with
$B(J,J')>0$, such that they both lie in the same positive cone of
$\Lambda$. Let $\underline J$ and $\underline J'$ be their
normalizations as before.
The arguments of theta function are $\tau\in \BH$, $\bfz\in \Lambda \otimes \mathbb{C}$ and $\bfmu\in \Lambda\otimes \mathbb{R}$. We let
$\bfb=\mathrm{Im}(\bfz)/\tau_2\in \Lambda \otimes \mathbb{R}$. In terms of this data, the indefinite theta function $\Theta_\bfmu^{JJ'}$ is defined as 
\be 
\label{indeftheta} 
\begin{split}
\Theta^{JJ'}_{\bfmu}\!(\tau,\bfz)=&\sum_{\bfk\in \Lambda+\bfmu} 
\tfrac{1}{2}\left( \sgn(B(\bfk+\bfb,J))-\sgn(B(\bfk+\bfb,J'))\right)\\
&(-1)^{B(\bfk,K)} q^{-\bfk^2/2}
e^{-2\pi i B(\bfz,\bfk)}.
\end{split} 
\ee  
One may show that the sum over $\Lambda$ is convergent \cite{ZwegersThesis}. However, $\Theta^{JJ'}_{\bfmu}$ does only transform as a modular form after  addition of certain non-holomorphic terms. References \cite{ZwegersThesis, MR2605321} explain that the modular completion $\widehat \Theta^{JJ'}_\bfmu$ of $\Theta^{JJ'}_\bfmu$ is obtained by substituting (rescaled) error functions for the sgn-functions in (\ref{indeftheta}). The completion $\widehat \Theta^{JJ'}_\bfmu$ then transforms as a modular form of weight $n/2$, and is explicitly given by
\be 
\label{hatTheta}
\begin{split}   
\widehat \Theta^{JJ'}_{\bfmu}\!(\tau,\bfz)=\sum_{\bfk\in \Lambda+\mu} &
\tfrac{1}{2}\left( E(\sqrt{2\tau_2}\,B(\bfk+\bfb, \underline
  J))-E(\sqrt{2\tau_2}\,B(\bfk+\bfb, \underline J'))\right)\\
& \times (-1)^{B(\bfk,K)} q^{-\bfk^2/2} e^{-2\pi i B(\bfz,\bfk)},
\end{split}  
\ee   
where $E(u):\mathbb{R}\to [-1,1]$ is a reparametrization of the error function,
\begin{equation}
E(u) = 2\int_0^u e^{-\pi t^2}dt = \text{Erf}(\sqrt{\pi}u).
\end{equation}
Note that in the limit $\tau_2\to \infty$, $E$ in (\ref{hatTheta})
approaches the original $\sgn$-function of (\ref{indeftheta}),
$$\lim_{\tau_2\to \infty} E\left(\sqrt{2\tau_2}\,u\right)=\sgn(u).$$
If we analytically continue $E$ to a function with complex argument,
then this limit is only convergent for
$-\frac{\pi}{4}<\mathrm{Arg}(u)<\frac{\pi}{4}$. 

The transformation properties under SL$(2,\BZ)$ follow from chapter 2 of Zwegers'
thesis \cite{ZwegersThesis} or Vign\'eras \cite{Vigneras:1977}. One
finds for the action of the generators on $\widehat
\Theta^{JJ'}_{\bfmu+K/2}(\tau,\bfz)$ 
\be
\label{theta_comp_mod}
\begin{split}
&\widehat \Theta^{JJ'}_{\bfmu+K/2}(\tau+1,\bfz)=e^{\pi i (\bfmu^2-K^2/4)}\, \widehat\Theta^{JJ'}_{\bfmu+K/2}(\tau,\bfz+\bfmu),\\
&\widehat\Theta^{JJ'}_{\bfmu+K/2}(-1/\tau,\bfz/\tau)=i(-i\tau)^{n/2} \exp\!\left(-\pi
  i \bfz^2/\tau+\pi i K^2/2\right) \widehat\Theta^{JJ'}_{K/2}(\tau,\bfz-\bfmu).
\end{split}
\ee

For our application, the  $\bar \tau$-derivative of $\widehat \Theta^{JJ'}_\bfmu$ is of particular
interest. This gives the ``shadow''\footnote{Since indefinite theta functions are mixed mock modular forms in general, the notion of ``shadow'' used here is slightly different from its definition for mock modular forms \cite{MR2605321}.} of
$\Theta^{JJ'}_\bfmu$, whose modular properties are easier to
determine than those of $\Theta^{JJ'}_\bfmu$. We obtain here 
\be
\label{shadow}
\begin{split} 
\partial_{\bar \tau} \widehat
\Theta^{JJ'}_\bfmu(\tau,\bfz)=& \Psi^J_\bfmu(\tau,\bfz)-\Psi^{J'}_\bfmu(\tau,\bfz),
\end{split}
\ee
with $\Psi^J_\bfmu$ (\ref{PsiJ}) the same function discussed in Appendix \ref{app_mod_forms}. The modular properties of $\Psi^J_\bfmu$ are  given in (\ref{Psi_trafos}), and can be obtained using standard Poisson resummation. 

The completion (\ref{hatTheta}) may simplify if the lattice $\Lambda$ contains vectors $\bfk_0\in \Lambda$ with norm $\bfk_0^2=0$. For such lattices $J$ and/or $J'$ can be chosen to equal such a vector, and careful analysis of the limit shows that the error function reduces to the original sgn-function \cite{ZwegersThesis}.  We assume now that $J'\in \Lambda$ such that $(J')^2=0$.  To ensure convergence of the sum, one needs to require furthermore that $B(\bfk+\bfb,J')\neq 0$ for any $\bfk\in \Lambda+K/2+\bfmu$, except if one also has $B(\bfk+\bfb,J)=0$. Then the completion $\widehat \Theta^{JJ'}_\bfmu$ is given by
\be  
\label{hat2Theta} 
\begin{split}   
\widehat \Theta^{JJ'}_{\bfmu}\!(\tau,\bfz)=\sum_{\bfk\in \Lambda+K/2+\mu} &
\tfrac{1}{2}\left( E(\sqrt{2\tau_2}B(\bfk+\bfb, \underline J))-\sgn(B(\bfk+\bfb, J'))\right)\\
& \times (-1)^{B(\bfk, K)} q^{-\bfk^2/2} e^{-2\pi i B(\bfz,\bfk)},
\end{split}  
\ee  
with shadow 
\be
\label{shadow2}
\partial_{\bar \tau}\Theta^{JJ'}_\bfmu(\tau,\bfz)=\Psi^J_\bfmu(\tau,\bfz).
\ee
We note that it is important here that $J'\in \Lambda$, since $\widehat \Theta^{JJ'}_{\bfmu}$ is otherwise not convergent. A divergent example is discussed in \cite[Appendix B.3]{Alexandrov:2017qhn}.

\section{Integrating over the fundamental domain}
\label{int_fund_dom}
In this appendix we discuss the recipe we use to evaluate the integral
over the $u$-plane. Let $\CF_Y$ be the compact set, whose
boundaries are given by the following arcs
\be  
\label{arcs}     
\begin{split}  
&1:\quad \tau=\tfrac{1}{2}+i\tau_2,\qquad \,\,\,\, \tau_2\in
[\tfrac{1}{2}\sqrt{3},Y],\\
& 2:\quad \tau=\tau_1+iY,\qquad  \,\,\,\, \tau_1\in [-\tfrac{1}{2},\tfrac{1}{2}],\\
&3:\quad  \tau=-\tfrac{1}{2}+i\tau_2,\qquad \tau_2\in
[\tfrac{1}{2}\sqrt{3},Y],\\
&4: \quad  \tau=i\,e^{i\varphi},\qquad \,\,\,\,\,\,\,\,\,\,\,\, \varphi\in [-\tfrac{\pi}{6},\tfrac{\pi}{6}].\\
\end{split}
\ee
We denote the non-compact subset obtained in the limit, $\lim_{Y\to
  \infty} \CF_Y$, by $\CF_{\infty}$, which can be chosen as the fundamental domain $\CF = \BH/$SL$(2,\BZ)$ of the modular group. We are interested in integrals of the form
\be
I_F=\lim_{Y\to\infty} \int_{\mathcal{F}_Y} d\tau\wedge d\bar \tau\, F,
\ee
where $F=F(\tau,\bar \tau)$ is a non-holomorphic function of $\tau$, which
transforms under $\operatorname{SL}(2,\BZ)$ as
\be
F\!\left(\frac{a\tau+b}{c\tau+d}, \frac{a\bar \tau+b}{c\bar
    \tau+d}\right)=|c\tau+d|^4 F(\tau,\bar \tau),
\ee
such that the integrand is modular invariant. Furthermore, we allow
that $F$ has a pole at $i\infty$ and its images in $\BQ$ under
SL$(2,\BZ)$, but assume that $F$ is regular elsewhere. We make furthermore the crucial
assumption that $F$ can be expressed as a total derivative
\be
\label{FG}
F=\frac{\partial H}{\partial \bar \tau},
\ee
where $H=H(\tau,\bar \tau)$ is a function which transforms as a weight two modular form. Note that $H$ is not unique since adding a weakly  holomorphic modular of weight 2 to $H$ does not change (\ref{FG}). Assuming that this ambiguity is fixed by other means, we see that the integrand is exact
\be
-d\left( H\, d\tau\right).
\ee
Therefore, by Stokes theorem $I_F$ equals
$$
I_F=-\lim_{Y\to \infty} \int_{\partial \mathcal{F}_Y}
H\, d\tau .
$$
Since the integrand is invariant under $\tau\to \tau+1$, the contribution of the arcs
(1) and (3) in (\ref{arcs}) add up to 0. Moreover, since the integrand
is invariant under $\tau\to -\frac{1}{\tau}$, the contribution due to
arc (4) vanishes. Thus what remains is arc (2), which is traversed in
counterclockwise direction. As a result we find
\be \label{thetaInt}
I_F=-\lim_{Y\to \infty} \int^{-\frac{1}{2}+iY}_{\frac{1}{2}+iY}
H\, d\tau =\left[H\right]_{q^0},
\ee
where the $\left[H\right]_{q^0}$ denotes the coefficient of $q^0$
of $H$.



\begin{thebibliography}{10}

\bibitem{Vafa:1994tf}
C.~Vafa and E.~Witten, \emph{{A Strong coupling test of S duality}},
  \href{http://dx.doi.org/10.1016/0550-3213(94)90097-3}{\emph{Nucl. Phys.} {\bf
  B431} (1994) 3--77}, [\href{https://arxiv.org/abs/hep-th/9408074}{{\tt
  hep-th/9408074}}].

\bibitem{Moore:1997pc}
G.~W. Moore and E.~Witten, \emph{{Integration over the u plane in Donaldson
  theory}}, {\emph{Adv. Theor. Math. Phys.} {\bf 1} (1997) 298--387},
  [\href{https://arxiv.org/abs/hep-th/9709193}{{\tt hep-th/9709193}}].

\bibitem{LoNeSha}
A.~Losev, N.~Nekrasov and S.~L. Shatashvili, \emph{{Issues in topological gauge
  theory}}, \href{http://dx.doi.org/10.1016/S0550-3213(98)00628-2}{\emph{Nucl.
  Phys.} {\bf B534} (1998) 549--611},
  [\href{https://arxiv.org/abs/hep-th/9711108}{{\tt hep-th/9711108}}].

\bibitem{Ne}
N.~A. Nekrasov, \emph{{Seiberg-Witten prepotential from instanton counting}},
  \href{http://dx.doi.org/10.4310/ATMP.2003.v7.n5.a4}{\emph{Adv. Theor. Math.
  Phys.} {\bf 7} (2003) 831--864},
  [\href{https://arxiv.org/abs/hep-th/0206161}{{\tt hep-th/0206161}}].

\bibitem{Pestun:2007rz}
V.~Pestun, \emph{{Localization of gauge theory on a four-sphere and
  supersymmetric Wilson loops}},
  \href{http://dx.doi.org/10.1007/s00220-012-1485-0}{\emph{Commun. Math. Phys.}
  {\bf 313} (2012) 71--129}, [\href{https://arxiv.org/abs/0712.2824}{{\tt
  0712.2824}}].

\bibitem{Kapustin:2009kz}
A.~Kapustin, B.~Willett and I.~Yaakov, \emph{{Exact Results for Wilson Loops in
  Superconformal Chern-Simons Theories with Matter}},
  \href{http://dx.doi.org/10.1007/JHEP03(2010)089}{\emph{JHEP} {\bf 03} (2010)
  089}, [\href{https://arxiv.org/abs/0909.4559}{{\tt 0909.4559}}].

\bibitem{Bershtein:2015xfa}
M.~Bershtein, G.~Bonelli, M.~Ronzani and A.~Tanzini, \emph{{Exact results for $
  \mathcal{N} $ = 2 supersymmetric gauge theories on compact toric manifolds
  and equivariant Donaldson invariants}},
  \href{http://dx.doi.org/10.1007/JHEP07(2016)023}{\emph{JHEP} {\bf 07} (2016)
  023}, [\href{https://arxiv.org/abs/1509.00267}{{\tt 1509.00267}}].

\bibitem{Witten:1988ze}
E.~Witten, \emph{{Topological Quantum Field Theory}},
  \href{http://dx.doi.org/10.1007/BF01223371}{\emph{Commun. Math. Phys.} {\bf
  117} (1988) 353}.

\bibitem{DONALDSON1990257}
S.~Donaldson, \emph{Polynomial invariants for smooth four-manifolds},
  \href{http://dx.doi.org/http://dx.doi.org/10.1016/0040-9383(90)90001-Z}{\emph{Topology}
  {\bf 29} (1990) 257 -- 315}.

\bibitem{Donaldson90}
S.~K. Donaldson and P.~B. Kronheimer, \emph{The geometry of four-manifolds /
  S.K. Donaldson and P.B. Kronheimer}.
\newblock Clarendon Press ; Oxford University Press Oxford : New York, 1990.

\bibitem{Witten:1994cg}
E.~Witten, \emph{{Monopoles and four manifolds}},
  \href{http://dx.doi.org/10.4310/MRL.1994.v1.n6.a13}{\emph{Math. Res. Lett.}
  {\bf 1} (1994) 769--796}, [\href{https://arxiv.org/abs/hep-th/9411102}{{\tt
  hep-th/9411102}}].

\bibitem{Witten:1995gf}
E.~Witten, \emph{{On S duality in Abelian gauge theory}},
  \href{http://dx.doi.org/10.1007/BF01671570}{\emph{Selecta Math.} {\bf 1}
  (1995) 383}, [\href{https://arxiv.org/abs/hep-th/9505186}{{\tt
  hep-th/9505186}}].

\bibitem{Gottsche:1996}
L.~G\"ottsche, \emph{Modular forms and donaldson invariants for 4-manifolds
  with $b_2^+= 1$}, {\emph{Journal of the American Mathematical Society} {\bf
  9} (1996) 827--843}, [\href{https://arxiv.org/abs/alg-geom/9506018}{{\tt
  alg-geom/9506018}}].

\bibitem{Gottsche:1996aoa}
L.~Gottsche and D.~Zagier, \emph{{Jacobi forms and the structure of Donaldson
  invariants for 4-manifolds with $b_2^+=1$}}, {\emph{{Selecta Math.}} {\bf 4}
  (1998) 69}.

\bibitem{LiQin1993}
W.~P. Li and Z.~Qin, \emph{Lower-degree donaldson polynomials of rational
  surfaces}, {\emph{J. Alg. Geom.} {\bf 2} (1993) 413--442}.

\bibitem{Kotschick1995}
D.~Kotschick and P.~Lisca, \emph{Instanton invariants of $\mathbb{P}^2$ via
  topology}, \href{http://dx.doi.org/10.1007/BF01460994}{\emph{Mathematische
  Annalen} {\bf 303} (1995) 345--371}.

\bibitem{ellingsrud1995wall}
G.~Ellingsrud and L.~G{\"o}ttsche, \emph{Wall-crossing formulas, bott residue
  formula and the donaldson invariants of rational surfaces}, {\emph{Quart. J.
  Math. Oxford Ser.} {\bf 49} (1998) 307--329},
  [\href{https://arxiv.org/abs/alg-geom/9506019}{{\tt alg-geom/9506019}}].

\bibitem{ZwegersThesis}
S.~P. Zwegers, \emph{Mock Theta Functions}.
\newblock PhD thesis, 2008.

\bibitem{MR2605321}
D.~Zagier, \emph{Ramanujan's mock theta functions and their applications (after
  {Z}wegers and {O}no-{B}ringmann)}, {\emph{Ast\'erisque} (2009) Exp. No. 986,
  vii--viii, 143--164 (2010)}.

\bibitem{Dixon:1990pc}
L.~J. Dixon, V.~Kaplunovsky and J.~Louis, \emph{{Moduli dependence of string
  loop corrections to gauge coupling constants}},
  \href{http://dx.doi.org/10.1016/0550-3213(91)90490-O}{\emph{Nucl. Phys.} {\bf
  B355} (1991) 649--688}.

\bibitem{Harvey:1995fq}
J.~A. Harvey and G.~W. Moore, \emph{{Algebras, BPS states, and strings}},
  \href{http://dx.doi.org/10.1016/0550-3213(95)00605-2}{\emph{Nucl. Phys.} {\bf
  B463} (1996) 315--368}, [\href{https://arxiv.org/abs/hep-th/9510182}{{\tt
  hep-th/9510182}}].

\bibitem{Borcherds:1996uda}
R.~E. Borcherds, \emph{Automorphic forms with singularities on
  {G}rassmannians}, {\emph{Invent.\ Math.\ {\textbf {132}} (1998) 491} (1998)
  }, [\href{https://arxiv.org/abs/alg-geom/9609022}{{\tt alg-geom/9609022}}].

\bibitem{zagier:1975}
D.~Zagier, \emph{Nombres de classes et formes modulaires de poids 3/2},
  {\emph{C.R. Acad. Sc. Paris} {\bf 281} (1975) 883}.

\bibitem{Malmendier:2008db}
A.~Malmendier and K.~Ono, \emph{{SO(3)-Donaldson invariants of $\mathbb{P}^2$
  and Mock Theta Functions}},
  \href{http://dx.doi.org/10.2140/gt.2012.16.1767}{\emph{Geom. Topol.} {\bf 16}
  (2012) 1767--1833}, [\href{https://arxiv.org/abs/0808.1442}{{\tt
  0808.1442}}].

\bibitem{Malmendier:2010ss}
A.~Malmendier, \emph{{Donaldson invariants of $\mathbb{P}^1 \times
  \mathbb{P}^1$ and Mock Theta Functions}},
  \href{http://dx.doi.org/10.4310/CNTP.2011.v5.n1.a5}{\emph{Commun. Num. Theor.
  Phys.} {\bf 5} (2011) 203--229}, [\href{https://arxiv.org/abs/1008.0175}{{\tt
  1008.0175}}].

\bibitem{Griffin:2012kw}
M.~Griffin, A.~Malmendier and K.~Ono, \emph{{SU(2)-Donaldson invariants of the
  complex projective plane}},
  \href{http://dx.doi.org/10.1515/forum-2013-6013}{\emph{Forum Math.} {\bf 27}
  (2015) 2003--2023}, [\href{https://arxiv.org/abs/1209.2743}{{\tt
  1209.2743}}].

\bibitem{Gaiotto:2009we}
D.~Gaiotto, \emph{{N=2 dualities}},
  \href{http://dx.doi.org/10.1007/JHEP08(2012)034}{\emph{JHEP} {\bf 08} (2012)
  034}, [\href{https://arxiv.org/abs/0904.2715}{{\tt 0904.2715}}].

\bibitem{Gaiotto:2009hg}
D.~Gaiotto, G.~W. Moore and A.~Neitzke, \emph{{Wall-crossing, Hitchin Systems,
  and the WKB Approximation}},  \href{https://arxiv.org/abs/0907.3987}{{\tt
  0907.3987}}.

\bibitem{Manschot:2007ha}
J.~Manschot and G.~W. Moore, \emph{{A Modern Farey Tail}},
  \href{http://dx.doi.org/10.4310/CNTP.2010.v4.n1.a3}{\emph{Commun. Num. Theor.
  Phys.} {\bf 4} (2010) 103--159}, [\href{https://arxiv.org/abs/0712.0573}{{\tt
  0712.0573}}].

\bibitem{Manschot:2009ia}
J.~Manschot, \emph{{Stability and duality in N=2 supergravity}},
  \href{http://dx.doi.org/10.1007/s00220-010-1104-x}{\emph{Commun. Math. Phys.}
  {\bf 299} (2010) 651--676}, [\href{https://arxiv.org/abs/0906.1767}{{\tt
  0906.1767}}].

\bibitem{Dabholkar:2012nd}
A.~Dabholkar, S.~Murthy and D.~Zagier, \emph{{Quantum Black Holes, Wall
  Crossing, and Mock Modular Forms}},
  \href{https://arxiv.org/abs/1208.4074}{{\tt 1208.4074}}.

\bibitem{Alexandrov:2016tnf}
S.~Alexandrov, S.~Banerjee, J.~Manschot and B.~Pioline, \emph{{Multiple
  D3-instantons and mock modular forms I}},
  \href{http://dx.doi.org/10.1007/s00220-016-2799-0}{\emph{Commun. Math. Phys.}
  {\bf 353} (2017) 379--411}, [\href{https://arxiv.org/abs/1605.05945}{{\tt
  1605.05945}}].

\bibitem{Cheng:2011ay}
M.~C.~N. Cheng and J.~F.~R. Duncan, \emph{{On Rademacher Sums, the Largest
  Mathieu Group, and the Holographic Modularity of Moonshine}},
  \href{http://dx.doi.org/10.4310/CNTP.2012.v6.n3.a4}{\emph{Commun. Num. Theor.
  Phys.} {\bf 6} (2012) 697--758}, [\href{https://arxiv.org/abs/1110.3859}{{\tt
  1110.3859}}].

\bibitem{Seiberg:1994rs}
N.~Seiberg and E.~Witten, \emph{{Electric - magnetic duality, monopole
  condensation, and confinement in N=2 supersymmetric Yang-Mills theory}},
  \href{http://dx.doi.org/10.1016/0550-3213(94)90124-4,
  10.1016/0550-3213(94)00449-8}{\emph{Nucl. Phys.} {\bf B426} (1994) 19--52},
  [\href{https://arxiv.org/abs/hep-th/9407087}{{\tt hep-th/9407087}}].

\bibitem{Seiberg:1994aj}
N.~Seiberg and E.~Witten, \emph{{Monopoles, duality and chiral symmetry
  breaking in N=2 supersymmetric QCD}},
  \href{http://dx.doi.org/10.1016/0550-3213(94)90214-3}{\emph{Nucl. Phys.} {\bf
  B431} (1994) 484--550}, [\href{https://arxiv.org/abs/hep-th/9408099}{{\tt
  hep-th/9408099}}].

\bibitem{AlvarezGaume}
L.~Alvarez-Gaume and S.~F. Hassan, \emph{{Introduction to S duality in N=2
  supersymmetric gauge theories: A Pedagogical review of the work of Seiberg
  and Witten}}, \href{http://dx.doi.org/10.1002/prop.2190450302}{\emph{Fortsch.
  Phys.} {\bf 45} (1997) 159--236},
  [\href{https://arxiv.org/abs/hep-th/9701069}{{\tt hep-th/9701069}}].

\bibitem{Bilal:1995hc}
A.~Bilal, \emph{{Duality in N=2 SUSY SU(2) Yang-Mills theory: A Pedagogical
  introduction to the work of Seiberg and Witten}},  in \emph{{Quantum fields
  and quantum space time. Proceedings, NATO Advanced Study Institute, Cargese,
  France, July 22-August 3, 1996}}, pp.~21--43, 1997.
\newblock \href{https://arxiv.org/abs/hep-th/9601007}{{\tt hep-th/9601007}}.

\bibitem{Shnir}
Y.~M. Shnir, \emph{{Magnetic Monopoles}}.
\newblock Springer-Verlag Berlin Heidelberg, 2005,
  \href{http://dx.doi.org/10.1007/3-540-29082-6}{10.1007/3-540-29082-6}.

\bibitem{Tern09}
J.~Terning, \emph{Modern Supersymmetry: Dynamics and Duality}.
\newblock Oxford Science Publications, 2009.

\bibitem{Tachikawa13}
Y.~Tachikawa, \emph{{N=2 supersymmetric dynamics for pedestrians}},
  {\emph{Lect. Notes Phys.} {\bf 890} (2014) },
  [\href{https://arxiv.org/abs/1312.2684}{{\tt 1312.2684}}].

\bibitem{Gaiotto:2014bja}
D.~Gaiotto, \emph{{Families of $\mathcal{N} =$ 2 Field Theories}},  in
  \emph{New Dualities of Supersymmetric Gauge Theories} (J.~Teschner, ed.),
  pp.~31--51.
\newblock 2016.
\newblock \href{https://arxiv.org/abs/1412.7118}{{\tt 1412.7118}}.
\newblock \href{http://dx.doi.org/10.1007/978-3-319-18769-3_2}{DOI}.

\bibitem{Nakajima:2003uh}
H.~Nakajima and K.~Yoshioka, \emph{{Lectures on instanton counting}},  in
  \emph{{CRM Workshop on Algebraic Structures and Moduli Spaces Montreal,
  Canada, July 14-20, 2003}}, 2003.
\newblock \href{https://arxiv.org/abs/math/0311058}{{\tt math/0311058}}.

\bibitem{Bouwknegt02}
S.~Wu, \emph{{The Geometry and Physics of the Seiberg-Witten equations}},  in
  \emph{Geometric Analysis and Applications to Quantum Field Theory}
  (P.~Bouwknegt and S.~Wu, eds.), pp.~31--51.
\newblock 2002.
\newblock \href{http://dx.doi.org/110.1007/978-1-4612-0067-3}{DOI}.

\bibitem{Matone:1995rx}
M.~Matone, \emph{{Instantons and recursion relations in N=2 SUSY gauge
  theory}}, \href{http://dx.doi.org/10.1016/0370-2693(95)00920-G}{\emph{Phys.
  Lett.} {\bf B357} (1995) 342--348},
  [\href{https://arxiv.org/abs/hep-th/9506102}{{\tt hep-th/9506102}}].

\bibitem{Witten90}
E.~Witten, \emph{{On the structure of topological phase of two-dimensional
  gravity}}, \href{http://dx.doi.org/10.1016/0550-3213(90)90449-N}{\emph{Nucl.
  Phys.} {\bf B340} (1990) 281--332}.

\bibitem{KMMN}
G.~Korpas, J.~Manschot, G.~W. Moore and I.~Nidaiev, \emph{{to appear}}, .

\bibitem{Labastida:2005zz}
J.~Labastida and M.~Marino, \emph{{Topological quantum field theory and four
  manifolds}}.
\newblock Springer, 2005.

\bibitem{Verlinde:1995mz}
E.~P. Verlinde, \emph{{Global aspects of electric - magnetic duality}},
  \href{http://dx.doi.org/10.1016/0550-3213(95)00431-Q}{\emph{Nucl. Phys.} {\bf
  B455} (1995) 211--228}, [\href{https://arxiv.org/abs/hep-th/9506011}{{\tt
  hep-th/9506011}}].

\bibitem{Petersson1950}
H.~Petersson, \emph{{Konstruktion der Modulformen und der zu gewissen
  Grenzkreisgruppen geh\"origen automorphen Formen von positiver reeller
  Dimension und die vollst\"andige Bestimmung ihrer Fourierkoeffizienten}},
  {\emph{S.-B. Heidelberger Akad. Wiss. Math.-Nat. Kl.} (1950) 417--494}.

\bibitem{Lerche:1988np}
W.~Lerche, A.~N. Schellekens and N.~P. Warner, \emph{{Lattices and Strings}},
  \href{http://dx.doi.org/10.1016/0370-1573(89)90077-X}{\emph{Phys. Rept.} {\bf
  177} (1989) 1}.

\bibitem{Barth}
C.~P. W.~Barth, K.~Hulek and A.~van~de Ven, \emph{Compact Complex Surfaces}.
\newblock Springer-Verlag Berlin Heidelberg, 2~ed., 2004.

\bibitem{Alexandrov:2017qhn}
S.~Alexandrov, S.~Banerjee, J.~Manschot and B.~Pioline, \emph{{Multiple
  D3-instantons and mock modular forms II}},
  \href{https://arxiv.org/abs/1702.05497}{{\tt 1702.05497}}.

\bibitem{Marino:1998bm}
M.~Marino and G.~W. Moore, \emph{{The Donaldson-Witten function for gauge
  groups of rank larger than one}},
  \href{http://dx.doi.org/10.1007/s002200050494}{\emph{Commun. Math. Phys.}
  {\bf 199} (1998) 25--69}, [\href{https://arxiv.org/abs/hep-th/9802185}{{\tt
  hep-th/9802185}}].

\bibitem{Klemm:1994qs}
A.~Klemm, W.~Lerche, S.~Yankielowicz and S.~Theisen, \emph{{Simple
  singularities and N=2 supersymmetric Yang-Mills theory}},
  \href{http://dx.doi.org/10.1016/0370-2693(94)01516-F}{\emph{Phys. Lett.} {\bf
  B344} (1995) 169--175}, [\href{https://arxiv.org/abs/hep-th/9411048}{{\tt
  hep-th/9411048}}].

\bibitem{Argyres:1994xh}
P.~C. Argyres and A.~E. Faraggi, \emph{{The vacuum structure and spectrum of
  N=2 supersymmetric SU(n) gauge theory}},
  \href{http://dx.doi.org/10.1103/PhysRevLett.74.3931}{\emph{Phys. Rev. Lett.}
  {\bf 74} (1995) 3931--3934},
  [\href{https://arxiv.org/abs/hep-th/9411057}{{\tt hep-th/9411057}}].

\bibitem{Douglas:1995nw}
M.~R. Douglas and S.~H. Shenker, \emph{{Dynamics of SU(N) supersymmetric gauge
  theory}}, \href{http://dx.doi.org/10.1016/0550-3213(95)00258-T}{\emph{Nucl.
  Phys.} {\bf B447} (1995) 271--296},
  [\href{https://arxiv.org/abs/hep-th/9503163}{{\tt hep-th/9503163}}].

\bibitem{Tachikawa:2011yr}
Y.~Tachikawa and S.~Terashima, \emph{{Seiberg-Witten Geometries Revisited}},
  \href{http://dx.doi.org/10.1007/JHEP09(2011)010}{\emph{JHEP} {\bf 09} (2011)
  010}, [\href{https://arxiv.org/abs/1108.2315}{{\tt 1108.2315}}].

\bibitem{Mochiziku:2009}
T.~Mochiziku, \emph{{Donaldson type invariants for algebraic surfaces}}.
\newblock Lecture Notes in Mathematics, no. 1972, Springer, 2009.

\bibitem{Marino:1998tb}
M.~Marino, G.~W. Moore and G.~Peradze, \emph{{Superconformal invariance and the
  geography of four manifolds}},
  \href{http://dx.doi.org/10.1007/s002200050694}{\emph{Commun. Math. Phys.}
  {\bf 205} (1999) 691--735}, [\href{https://arxiv.org/abs/hep-th/9812055}{{\tt
  hep-th/9812055}}].

\bibitem{Serre}
J.~P. Serre, \emph{A course in arithmetic}.
\newblock Graduate Texts in Mathematics, no. 7, Springer, New York, 1973.

\bibitem{Zagier92}
D.~Zagier, \emph{Introduction to modular forms; From Number Theory to Physics}.
\newblock Springer, Berlin (1992), pp. 238-291, 1992.

\bibitem{Bruinier08}
G.~H. J.H.~Bruinier, G. van der~Geer and D.~Zagier, \emph{The 1-2-3 of Modular
  Forms}.
\newblock Springer-Verlag Berlin Heidelberg, 2008,
  \href{http://dx.doi.org/10.1007/978-3-540-74119-0}{10.1007/978-3-540-74119-0}.

\bibitem{Vigneras:1977}
M.-F. Vign\'eras, \emph{{S\'eries th\^eta des formes quadratiques
  ind\'efinies}}, {\emph{Springer Lecture Notes} {\bf 627} (1977) 227 -- 239}.

\end{thebibliography}

\providecommand{\href}[2]{#2}\begingroup\raggedright\endgroup

\end{document}